\documentclass[twocolumn]{aastex62}

\graphicspath{{./}{figures/}}

\received{11th June, 2019}
\revised{20th August, 2019}
\accepted{19th September, 2019}

\submitjournal{ApJ}

\shortauthors{Bennet et al.}

\begin{document}

\title{The M101 Satellite Luminosity Function and the Halo to Halo Scatter Among  Local Volume Hosts}

\correspondingauthor{Paul Bennet}
\email{paul.bennet@ttu.edu}

\author{P. Bennet}
\affiliation{Physics \& Astronomy Department, Texas Tech University, Box 41051, Lubbock, TX 79409-1051, USA}
\author[0000-0003-4102-380X]{D. J. Sand}
\affiliation{Steward Observatory, University of Arizona, 933 North Cherry Avenue, Rm. N204, Tucson, AZ 85721-0065, USA}
\author{D. Crnojevi\'c}
\affiliation{University of Tampa, 401 West Kennedy Boulevard, Tampa, FL 33606, USA}
\affiliation{Physics \& Astronomy Department, Texas Tech University, Box 41051, Lubbock, TX 79409-1051, USA}
\author{K. Spekkens}
\affiliation{Department of Physics and Space Science, Royal Military College of Canada P.O. Box 17000, Station Forces Kingston, ON K7K 7B4, Canada}
\affiliation{Department of Physics, Engineering Physics and Astronomy, Queen’s University, Kingston, ON K7L 3N6, Canada}
\author{A. Karunakaran}
\affiliation{Department of Physics, Engineering Physics and Astronomy, Queen’s University, Kingston, ON K7L 3N6, Canada}
\author{D. Zaritsky}
\affiliation{Steward Observatory, University of Arizona, 933 North Cherry Avenue, Rm. N204, Tucson, AZ 85721-0065, USA}
\author{B. Mutlu-Pakdil}
\affiliation{Steward Observatory, University of Arizona, 933 North Cherry Avenue, Rm. N204, Tucson, AZ 85721-0065, USA}

\begin{abstract}

We have obtained deep Hubble Space Telescope (HST) imaging of 19 dwarf galaxy candidates in the vicinity of M101. 
Advanced Camera for Surveys (ACS) HST photometry for 2 of these objects showed resolved stellar populations and Tip of the Red Giant Branch (TRGB) derived distances (D$\sim$7 Mpc) consistent with M101 group membership.  
The remaining 17 were found to have no resolved stellar populations, meaning they are either part of the background NGC 5485 group or are distant low surface brightness (LSB) galaxies. 
It is noteworthy that many LSB objects which had previously been assumed to be M101 group members based on projection have been shown to be background objects, indicating the need for future diffuse dwarf surveys to be very careful in drawing conclusions about group membership without robust distance estimates. 
In this work we update the satellite luminosity function (LF) of M101 based on the presence of these new objects down to M$_V$=$-$8.2. M101 is a sparsely populated system with only 9 satellites down to M$_V$$\approx$$-$8, as compared to 26 for M31 and 24.5$\pm$7.7 for the median host in the Local Volume.
This makes M101 by far the sparsest group probed to this depth, though M94 is even sparser to the depth it has been examined (M$_V$=$-$9.1).  
M101 and M94 share several properties that mark them as unusual compared to the other Local Volume galaxies examined: they have a very sparse satellite population but also have high star forming fractions among these satellites; such properties are also found in the galaxies examined as part of the SAGA survey. We suggest that these properties appear to be tied to the wider galactic environment, with more isolated galaxies showing sparse satellite populations which are more likely to have had recent star formation, while those in dense environments have more satellites which tend to have no ongoing star formation.
Overall our results show a level of halo-to-halo scatter between galaxies of similar mass that is larger than is predicted in the $\Lambda$CDM model. 
\end{abstract}


\section{Introduction} \label{sec:intro}

Observations on large scales ($\gtrsim$10 Mpc) are consistent with a Universe dominated by dark energy and cold dark matter, along with a small contribution from baryons -- the so-called $\Lambda$ Cold Dark Matter ($\Lambda$CDM) model for structure formation \citep[e.g.][]{Planck16}.  Despite this success, challenges remain on smaller, subgalactic scales where straightforward expectations for the faint end of the galaxy luminosity function (LF) are not met \citep[see][for a recent review]{Bullock17} including the `missing satellites problem' \citep[e.g.][]{moore99,klypin99}, `too big to fail' \citep[e.g.][]{Boylan11,boylan12} and the apparent planes of satellites around nearby galaxies \citep[e.g.][]{Pawlowski12,ibata13,Muller18}.  

Significant progress has been made in reconciling these small-scale $\Lambda$CDM `problems' on both the theoretical \citep{Brooks13,Sawala16,Wetzel16} and observational \citep[e.g. ][most recently around the Milky Way, MW]{Torrealba18,Torrealba18b,Koposov18} fronts, although the focus has been on the Local Group and its satellite system.  Ultimately, to fully test the $\Lambda$CDM model for structure formation, studies beyond the Local Group are necessary in order to sample primary halos with a range of masses, morphologies and environments.  This work is now beginning in earnest using wide-field imaging datasets, as well as spectroscopy, centered around primary galaxies with a range of masses  \citep[e.g.][]{chiboucas13,Sand14,Sand15a,crnojevic14b,Muller15,crnojevic16,carlin16,toloba16,bennet17,Danieli17,Smercina17,Geha17,Smercina18,Crnojevic19}.  
Field searches are also uncovering a plethora of faint dwarf galaxy systems using a variety of techniques \citep[e.g.][]{Tollerud15,Sand15b,Leisman17,Greco18,Bennet18,Zaritsky19}.

One opportunity is to measure the dispersion in substructure properties among Milky Way-like halos, partially to understand if the Local Group has unusual substructure properties, and to help guide simulations which are addressing $\Lambda$CDM's so-called problems.  Initial results in this arena are exciting -- the Satellites Around Galactic Analogs \citep[SAGA;][]{Geha17} survey has found that the halo to halo scatter in bright satellite numbers is higher than expected from abundance matching expectations.  SAGA also found many examples of star forming dwarf satellites, in contrast with the dwarf population in the Local Group.  Additionally, a recent search for faint satellites around M94 ($D$=4.2 Mpc), another Milky Way analogue, found only two satellites with $M_\star$$>$4$\times$10$^5$ $M_{\odot}$ in comparison to the eight systems found around the Milky Way \citep{Smercina18}\footnote Note there are also 14 additional dwarf candidates with velocities consistent with M94 \citep{karachentsev13}, however these objects were outside of of the search radius in \cite{Smercina18}. 
At the bright end of the satellite LF the number of objects is small and therefore the statistical power is low (e.g. the 8 satellites around the MW is $\sim$3$\sigma$ discrepant from zero), this provides additional motivation to explore the faint end of the LF where the numbers of satellites is larger and therefore produce more robust statistics. 
Despite the large observational effort, the number of Milky Way-like systems studied is still small, and further work is needed to quantify the observed range in substructure properties.

Here we present {\it Hubble Space Telescope} follow-up to 19 dwarf galaxy candidates recently discovered around M101  \citep[to which we assume a distance of $D$=7 Mpc throughout this work;][]{Lee12,tikhonov15}, both to determine their membership status and to construct a satellite LF.  M101 is an excellent system for comparing with our own Local Group, as its stellar mass \citep[$\sim$5.3$\times$10$^{10}$ M$_{\odot}$;][]{vanDokkum14} is similar to that of the MW to within the uncertainties \citep[e.g.][]{mcmillan11}.  M101 also has an `anemic' low mass stellar halo \citep{vanDokkum14} that nonetheless shows signs of past galaxy interactions \citep[e.g.][]{mihos13}.  Measuring the diversity of satellite populations around MW-like systems is a main driver for this work.

An outline of the paper follows.  In Section~\ref{sec:history}, we give context and an overview of recent dwarf galaxy searches around M101, and how our 19 dwarf targets were selected for follow-up.   In Section~\ref{sec:data} we describe the {\it HST} photometry and artificial star tests. In Section~\ref{sec:results} we present  the properties of the dwarf populations around M101; we also discuss the statistical properties of the population of M101 dwarf candidates that were not observed by {\it HST}. Next, in Section~\ref{sec:disc} we discuss the luminosity function of the M101 system and compare it to other nearby  Local Volume galaxies and those found in the SAGA survey.  We also compare the dwarf star forming fractions within these groups and explore a potential correlation between host environment and star formation fraction within the satellites.   Finally, we summarize and conclude in Section~\ref{sec:conclusion}.

\section{Dwarf Candidates Around M~101} \label{sec:history}

Although traditionally thought to have a relatively poor satellite galaxy population \citep[e.g.][]{bremnes99}, dwarf galaxy searches around M101 have been reinvigorated by the surge in interest in the low surface brightness (LSB) universe.  Using a set of specially designed telephoto lenses, the Dragonfly team identified seven new diffuse dwarf galaxy candidates in a $\sim$9 deg$^2$ region around M101 \citep{merritt14}.  Out of these seven dwarf candidates, three were identified as true M101 dwarfs based on their {\it HST}-derived tip of the red giant branch (TRGB) distance \citep[M101 DF1, M101 DF2, and M101 DF3;][]{Danieli17}; the remaining four were found to be background sources, perhaps associated with the elliptical galaxy NGC5485, at $D$$\sim$27 Mpc \citep{Merritt16}.
Other small telescope searches have also identified M101 dwarf candidates in recent years, with \citet{karachentsev15} reporting on four additional objects \citep[DwA through DwD; see also][]{Java16}.   A search based on Sloan Digital Sky Survey imaging identified six additional dwarf galaxy candidates around M101 \citep{muller17}, although these objects were beyond the nominal virial radius of M101 in projection \citep[$\sim$260 kpc;][]{merritt14}.

Following on from these dwarf searches, \citet{bennet17} used data from the Wide portion of the Canada France Hawaii Telescope Legacy Survey (CFHTLS) to perform a semi-automated search for dwarfs in a $\sim$9 deg$^2$ region around M101, both to compare with previous work, and to develop a robust algorithm which could then be applied to even larger wide-field public imaging datasets.  This search found all of the previously identified dwarf candidates within its footprint, along with 39 additional dwarf candidates. One key aspect of this CFHTLS semi-automated search is that it conducted extensive simulations with fake dwarf galaxies implanted into the data, thus providing well-defined dwarf galaxy completeness limits.  It is from this set of candidate dwarfs that the 19 targets in the current work are drawn from.

We note that a single dwarf candidate in the CFHTLS sample, Dw26, was reported to have an HI detection at $v_{sys}$=11,000 km s$^{-1}$, indicating that it is a background galaxy at $D$$\approx$150 Mpc.  Other than this object, no further distance information was reported in \citet{bennet17}, making any conclusions about M101's satellite LF difficult.

Some distance information for the dwarf candidates around M101 can be gleaned through the technique of surface brightness fluctuations \citep[SBF; ][]{Tonry88,Tonry01,Cantiello18,carlsten19a}.  
Using a new calibration of the SBF technique based on TRGB distances, \citet{carlsten19} found that two of the dwarf candidates in \citet{bennet17} are likely satellites of M101 (DwA and Dw9), while two others (Dw15 and Dw21) are promising targets for follow-up. This SBF analysis of the M101 dwarfs used the same CFHTLS data that was used for the original detection in \cite{bennet17}. 

\begin{figure*}
 \begin{center}
 \includegraphics[width=16cm]{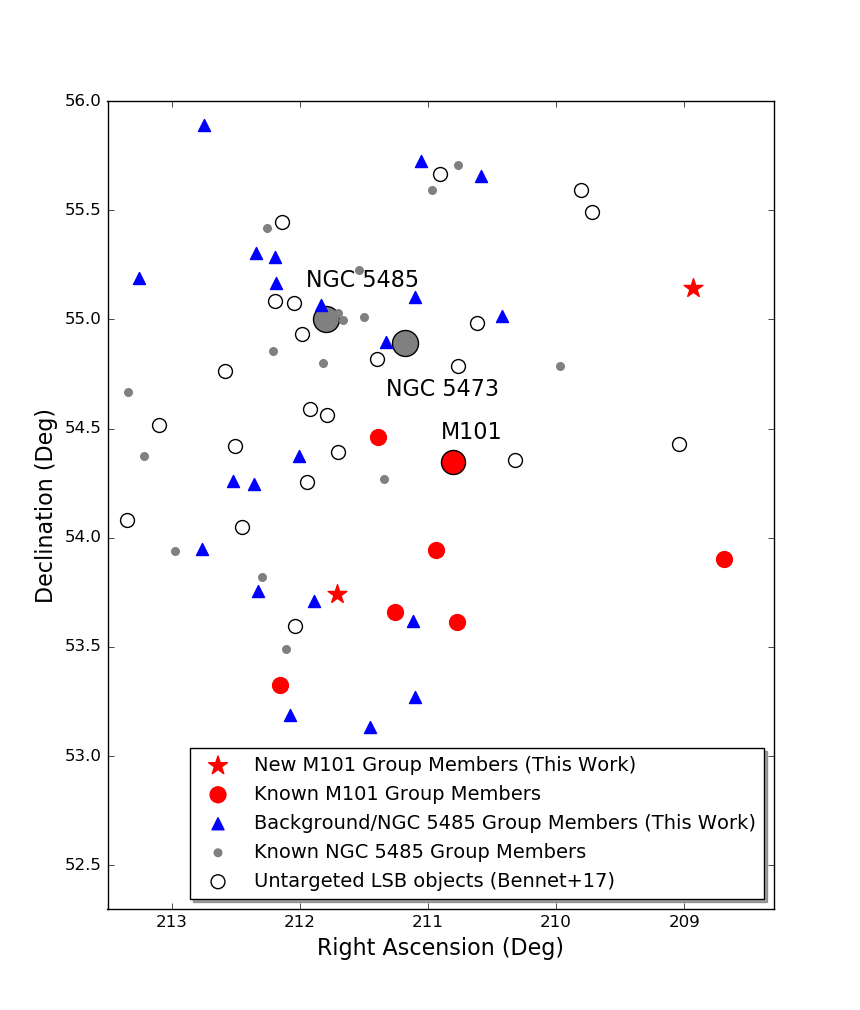}
 \caption{Spatial map of the M101 group and its surrounding region, as well as the background NGC~5485 group, which is nearby in projection. M101 satellites are shown in red, with dots showing previously confirmed members and stars the newly confirmed members; M101 itself is labelled and shown by the large red dot. NGC 5485 group members are shown by grey dots, NGC 5485 and NGC 5473 (the secondary, large member of the NGC~5473 group) are labelled and shown by the large dots. Unresolved objects as seen in the {\it HST} data in this work are shown by blue triangles; these may be NGC 5485 group members or background galaxies. The LSB objects that were not targeted by {\it HST} are shown by hollow circles. North is up, east is left. \label{fig:pos}}
 \end{center}
\end{figure*}

In the current work we present {\it HST} results for 19 of the dwarf candidates reported by \citet{bennet17}, four of which were first identified by \citet{karachentsev15}.  We list these objects in Table~\ref{tab:obj_res} and \ref{tab:obj_unres}.  As we discuss below, these 19 dwarfs are a representative sample of the entire diffuse dwarf candidate population around M101.  After determining the membership status of these dwarf candidates, we construct the satellite LF for M101, and compare it with other MW analogues to get an initial measure of the halo to halo scatter in this population.  A plot of the spatial distribution of M101 dwarfs and dwarf candidates are shown in Figure~\ref{fig:pos} for reference.

\section{HST Data and Photometry} \label{sec:data}

\floattable
\begin{deluxetable}{c|c|c}
\tablecaption{Confirmed M101 Dwarfs \label{tab:obj_res}}
\tablehead{
\colhead{Name} & \colhead{DwA} & \colhead{Dw9}}

\startdata
RA (J2000) & 14:06:49.9$\pm$1.0'' & 13:55:44.8$\pm$3.0'' \\
Dec (J2000) & +53:44:29.8$\pm$0.8'' & +55:08:45.6$\pm$2.1''\\
m$_{V}$ (CFHTLS) (mag) & 19.2$\pm$0.1 & 20.8$\pm$0.1 \\
m$_{V}$ (HST) (mag) & 19.6$\pm$0.2 & 21.0$\pm$0.2 \\
M$_{V}$ (HST) (mag) & $-$9.5$\pm$0.2 & $-$8.2$\pm$0.2 \\
r$_{h}$ (CFHTLS) (arcsec) & 10.92$\pm$0.23 & 7.66$\pm$0.64 \\
r$_{h}$ (HST) (arcsec) & 12.6$\pm$1.2 & 10.8$\pm$2.4 \\
r$_{h}$ (HST) (pc) & 417$\pm$40 & 384$\pm$85 \\
$\mu$(V,0) (mag arcsec$^{-2}$) & 26.0$\pm$0.3 & 27.2$\pm$0.5 \\
Mass (M$_{\odot}$) & 7.0$\pm$0.4$\times$10$^{5}$ & 2.0$\pm$0.1$\times$10$^{5}$ \\
Ellipticity & 0.33$\pm$0.06 & $\leq$0.37 \\
Distance (Mpc) & 6.83$^{+0.27}_{-0.26}$ & 7.34$^{+0.39}_{-0.37}$\\
Projected Distance from M101 (kpc) & 100 & 160 \\
\enddata
\tablenotetext{}{ Properties derived using a fixed exponential profile}
\end{deluxetable}

\floattable
\begin{deluxetable}{c|cc|cccc|cc}
\tablecaption{Unresolved Dwarf candidates \label{tab:obj_unres}}
\tablehead{
\colhead{Name} & \colhead{RA} & \colhead{Dec} & \colhead{V-band} & \colhead{V-band} & \colhead{F606W} & \colhead{F814W} & \colhead{Half light} & \colhead{Half light}\\
\colhead{} & \colhead{} & \colhead{} & \colhead{Magnitude} & \colhead{Magnitude} &\colhead{Magnitude} & \colhead{Magnitude} & \colhead{radius (CFHTLS)} & \colhead{radius (HST)}\\
\colhead{} & \colhead{} & \colhead{} & \colhead{(CFHTLS)} & \colhead{(HST)} & \colhead{(HST)} &\colhead{(HST)} & \colhead{(arcsec)} & \colhead{(arcsec)\tablenotemark{a}}}
\colnumbers
\startdata
DwB\tablenotemark{b} & 14:08:43.7 & +55:10:02 & 20.3$\pm$0.1 & 20.7$\pm$0.7 & 20.5$\pm$0.7 & 19.6$\pm$0.6 & 6.95$\pm$0.54 & 6.2$\pm$1.6 \\
DwC & 14:05:18.2 & +54:53:52 & 20.2$\pm$0.2 & 20.9$\pm$0.8 & 20.8$\pm$0.8 & 20.2$\pm$0.8 & 7.90$\pm$1.6 & 5.1$\pm$2.7 \\
DwD & 14:04:24.8 & +53:16:11 & 19.3$\pm$0.1 & 19.6$\pm$0.6 & 19.4$\pm$0.6 & 19.8$\pm$0.2 & 9.16$\pm$0.47 & 8.8$\pm$1.8 \\
Dw1 & 14:10:59.7 & +55:53:29 & 20.5$\pm$0.3 & 20.0$\pm$0.4 & 19.8$\pm$0.4 & 19.2$\pm$0.5 & 10.3$\pm$6.6 & 10.9$\pm$2.1 \\
Dw2 & 14:09:22.0 & +55:18:14 & 20.5$\pm$0.1 & 20.6$\pm$0.4 & 20.4$\pm$0.5 & 20.3$\pm$0.6 & 5.52$\pm$0.39 & 5.8$\pm$2.1\\
Dw3 & 14:08:45.8 & +55:17:14 & 19.5$\pm$0.1 & 19.7$\pm$0.5 & 19.5$\pm$0.5 & 19.1$\pm$0.5 & 7.11$\pm$0.42 & 5.7$\pm$1.8\\
Dw4 & 14:13:01.7 & +55:11:16 & 20.0$\pm$0.1 & 20.1$\pm$0.1 & 19.9$\pm$1.0 & 19.8$\pm$0.4 & 6.88$\pm$0.33 & 7.4$\pm$0.8\\
Dw5 & 14:04:13.0 & +55:43:34 & 20.1$\pm$0.2 & 20.6$\pm$0.5 & 20.4$\pm$0.5 & 20.0$\pm$0.6 & 7.64$\pm$0.86 & 3.7$\pm$1.1\\
Dw6 & 14:02:20.1 & +55:39:17 & 19.6$\pm$0.1 & 19.6$\pm$0.4 & 19.4$\pm$0.4 & 18.8$\pm$0.5 & 8.34$\pm$0.37 & 7.4$\pm$1.3\\
Dw7 & 14:07:21.0 & +55:03:51 & 20.1$\pm$0.1 & 21.2$\pm$0.3 & 21.0$\pm$0.3 & 20.6$\pm$0.3 & 4.70$\pm$0.20 & 3.9$\pm$0.5\\
Dw8 & 14:04:24.9 & +55:06:13 & 19.5$\pm$0.1 & 19.6$\pm$0.4 & 19.4$\pm$0.4 & 18.9$\pm$0.5 & 5.70$\pm$0.20 & 6.2$\pm$1.6\\
Dw10 & 14:01:40.4 & +55:00:57 & 22.0$\pm$0.2 & 22.1$\pm$0.6 & 21.9$\pm$0.6 & 21.7$\pm$0.9 & 4.63$\pm$0.97 & 3.2$\pm$1.2\\
Dw11 & 14:10:04.8 & +54:15:29 & 20.4$\pm$0.1 & 20.2$\pm$0.8 & 20.4$\pm$0.8 & 19.8$\pm$0.7 & 4.26$\pm$0.19 & 5.5$\pm$0.7\\
Dw12 & 14:09:26.0 & +54:14:51 & 20.5$\pm$0.1 & 20.8$\pm$0.4 & 21.0$\pm$0.4 & 19.9$\pm$0.3 & 4.32$\pm$0.25 & 2.9$\pm$0.3\\
Dw13 & 14:08:01.2 & +54:22:30 & 20.0$\pm$0.1 & 19.9$\pm$0.6 & 20.1$\pm$0.6 & 19.3$\pm$0.4 & 3.91$\pm$0.10 & 3.6$\pm$0.5\\
Dw14 & 14:11:03.2 & +53:56:50 & 21.2$\pm$0.1 & 20.4$\pm$0.8 & 20.6$\pm$0.8 & 19.9$\pm$1.0 & 5.76$\pm$0.30 & 6.2$\pm$1.3\\
Dw15 & 14:09:17.5 & +53:45:30 & 20.6$\pm$0.4 & 20.9$\pm$0.5 & 21.1$\pm$0.5 & 20.4$\pm$0.9 & 8.80$\pm$6.70 & 8.0$\pm$2.0\\
DF4 & 14:07:33.4 & +54:42:36 & 20.1$\pm$0.1 & 19.9$\pm$0.1 & 19.7$\pm$0.1 & 19.4$\pm$0.2 & 17.9$\pm$0.9 & 16.7$\pm$1.7 \\
DF5 & 14:04:28.1 & +55:37:00 & 20.7$\pm$0.2 & 20.8$\pm$0.2 & 20.6$\pm$0.2 & 19.9$\pm$0.3 & 10.8$\pm$2.6 & 8.9$\pm$0.9 \\
DF6 & 14:08:19.0 & +55:11:24 & 21.0$\pm$0.2 &  21.0$\pm$0.3\tablenotemark{c} &  20.8$\pm$0.3\tablenotemark{c} &  21.2$\pm$0.3\tablenotemark{c} & 22.8$\pm$2.1 &  22.8$\pm$2.8\tablenotemark{c} \\
DF7 & 14:05:48.3 & +55:07:58 & 21.0$\pm$0.4 & 20.6$\pm$0.3 & 20.4$\pm$0.3 & 19.2$\pm$0.3 & 28.0$\pm$14.0 & 12.2$\pm$1.8 \\
\enddata
\tablenotetext{a}{Derived from F606W images}
\tablenotetext{b}{Spectroscopically confirmed to be a member of the NGC 5485 group by HI observations (see Karunakaran et al. in prep)}
\tablenotetext{c}{From aperture photometry}
\end{deluxetable}

We obtained {\it HST} images (GO-14796; PI: Crnojevi\'c) of 19 of the dwarf candidates around M101 that were found as part of \cite{bennet17}. This {\it HST} follow-up was obtained via the Wide Field Camera (WFC) of the Advanced Camera for Surveys (ACS). Each target was observed for one orbit split evenly between the F606W and F814W filters (exposure time of $\sim$1200 seconds per filter). We did not dither to fill in the ACS chip gap, as our dwarf candidates easily fit onto one of the chips. 

We perform PSF-fitting photometry on the provided {\it .flt} images using the DOLPHOT v2.0 photometric package (with the ACS module), a modified version of HSTphot \citep{dolphin00}. For this work we use the suggested input parameters from the DOLPHOT's User Guide\footnote{http://americano.dolphinsim.com/dolphot/dolphotACS.pdf}, including corrections for charge transfer efficiency losses. Quality cuts are then applied using the following criteria: the derived photometric errors must be $\leq$0.3 mag in both bands, the sum of the crowding parameter in both bands is $\leq$1 and the squared sum of the sharpness parameter is $\leq$0.075. Detailed descriptions of these parameters can be found in \cite{dolphin00}.

We also performed artificial star tests to assess our photometric errors and incompleteness in the {\it HST} data. For these tests, artificial stars are distributed evenly across the image and in color-magnitude parameter space, extending up to 2 magnitudes below the faintest stars in the original CMD (after quality cuts) to account for objects that may have been up-scattered by noise. For each image we inject a total of 10 times the number of stars detected in the real data, after quality cuts, ensuring useful statistics. These artificial stars are injected one at a time by DOLPHOT so as not to induce crowding. Then quality cuts are used with the same criteria as the original image. 
These tests found the 50\% completeness limit for $F814W$ to be $\approx$26.9 mag and for $F606W$ to be $\approx$27.5 mag  and this was found to be consistent across all {\it HST} images.

We correct the derived magnitudes for foreground extinction on a star-by-star basis using the \cite{Schlafly11} calibration of the \cite{schlege98} dust maps.

\section{The Dwarf Candidate Population} \label{sec:results}

Of the 19 objects projected around M101 that were selected for {\it HST} follow-up, we see two distinct populations. The first group resolves into stars as is expected for a member of the M101 group (D$\sim$7 Mpc), and consists of two targets, DwA and Dw9. 
Meanwhile the other 17 targets appear as unresolved diffuse emission, indicating that the TRGB is too faint to be detected in our {\it HST} imaging, and that they are in the background. Beyond this, there are a further 23 diffuse dwarfs from \citet{bennet17} that were not imaged with {\it HST}, and we statistically assess their M101 membership status in Section~\ref{subsec:obj_not_tar}.  The spatial distribution of all of the resolved and diffuse dwarfs around M101 are shown in Figure~\ref{fig:pos}.

\begin{figure*}
 \begin{center}
 \includegraphics[width=7.5cm]{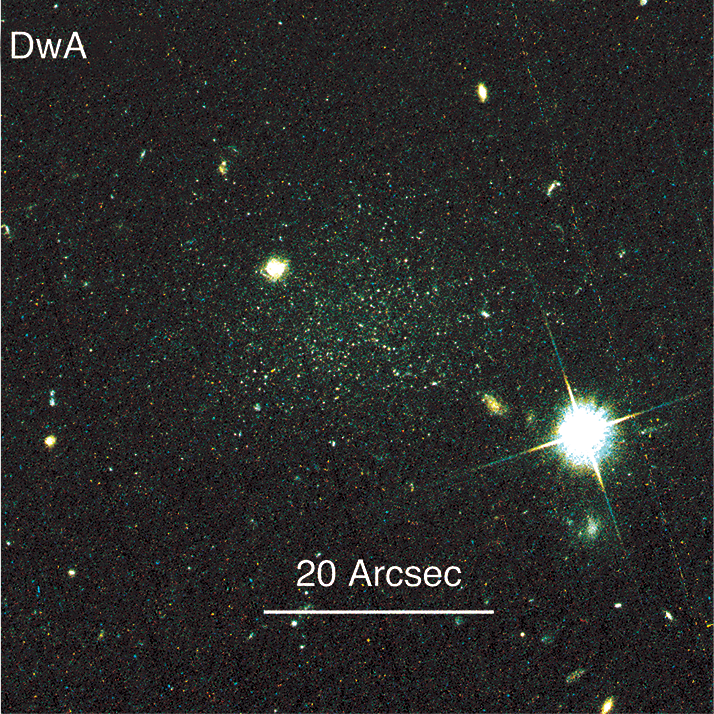}
  \includegraphics[width=7.45cm]{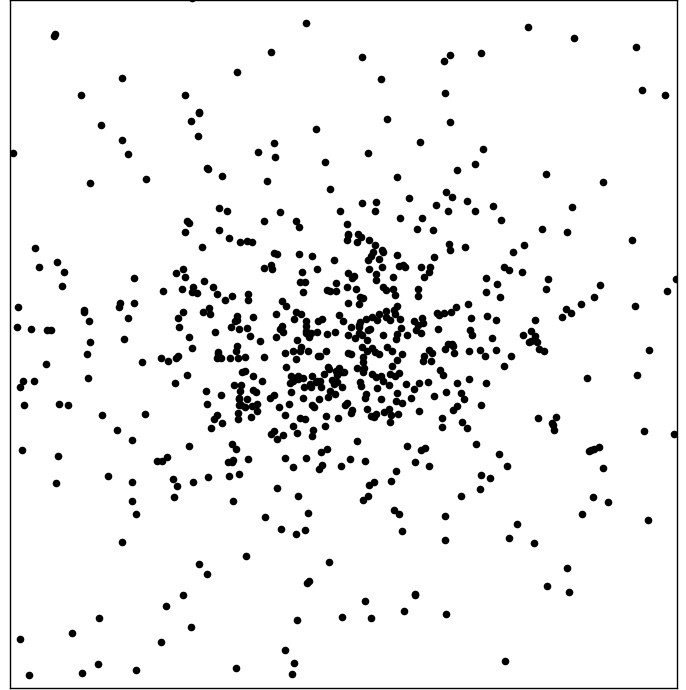}
 \includegraphics[width=7.5cm]{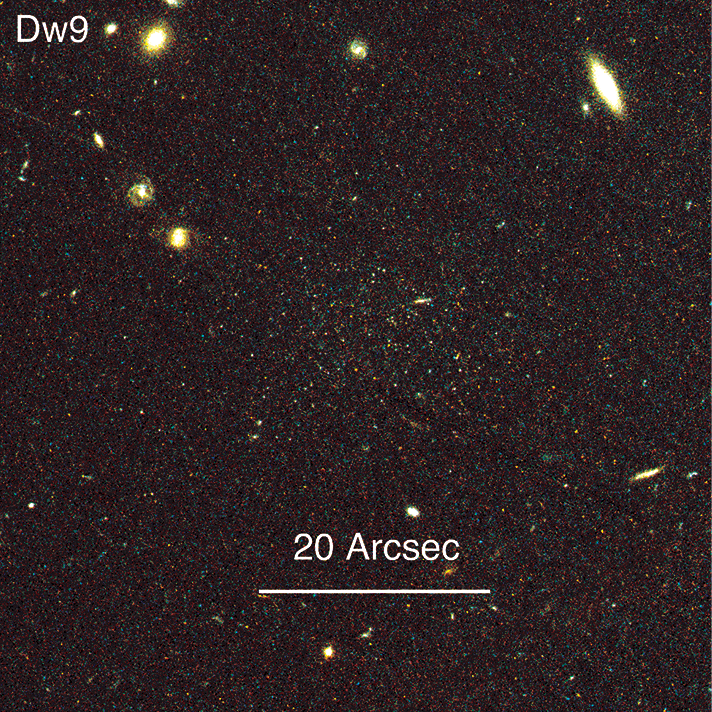}
  \includegraphics[width=7.45cm]{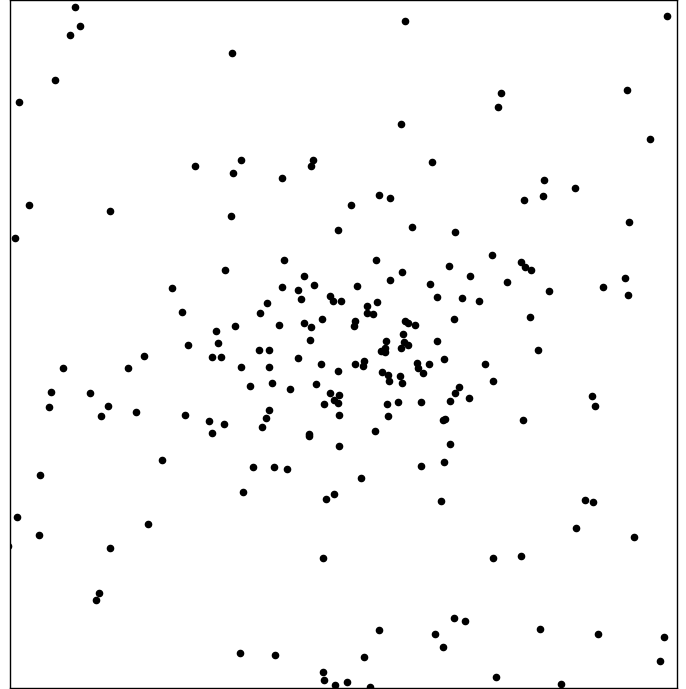}
 \caption{Left: Colorized image cutouts of the resolved dwarf candidates from {\it HST/ACS}, DwA and Dw9.  Right: Plots of all the point sources identified by DOLPHOT after quality cuts. 
Images are 1.0'x1.0', north is up, east is left. DwA and Dw9 contrast with the other dwarfs in our sample which have no overdensity of stellar objects at their position; see Figure~\ref{fig:pic_unresolved}.\label{fig:pic_resolved}}
 \end{center}
\end{figure*}

\begin{figure*}
 \begin{center}
 \includegraphics[width=7.5cm]{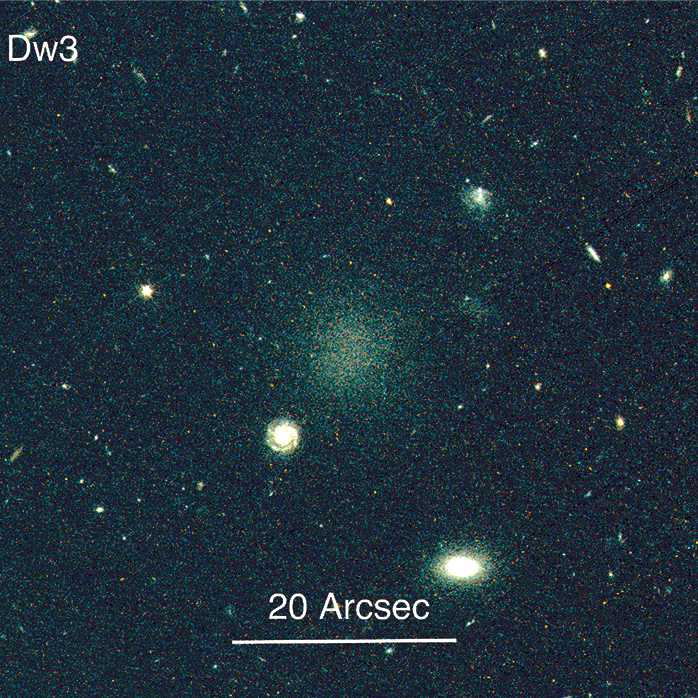}
  \includegraphics[width=7.45cm]{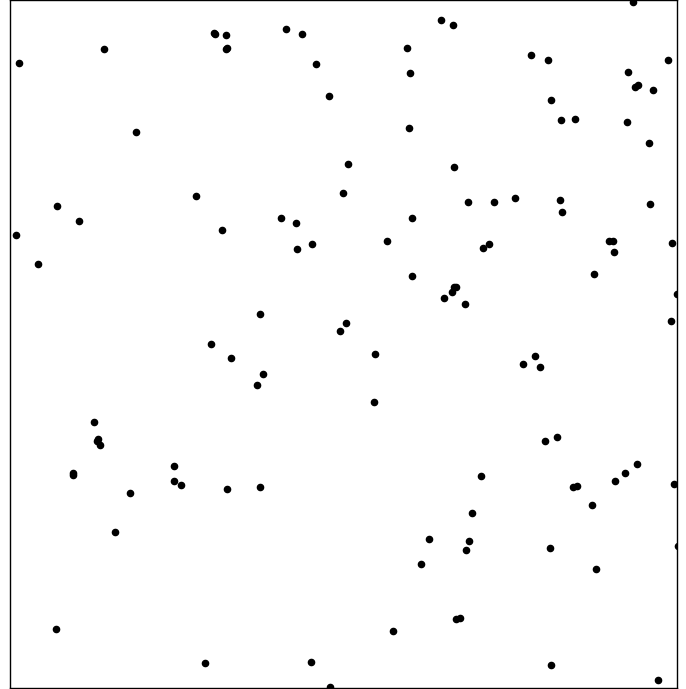}
 \includegraphics[width=7.5cm]{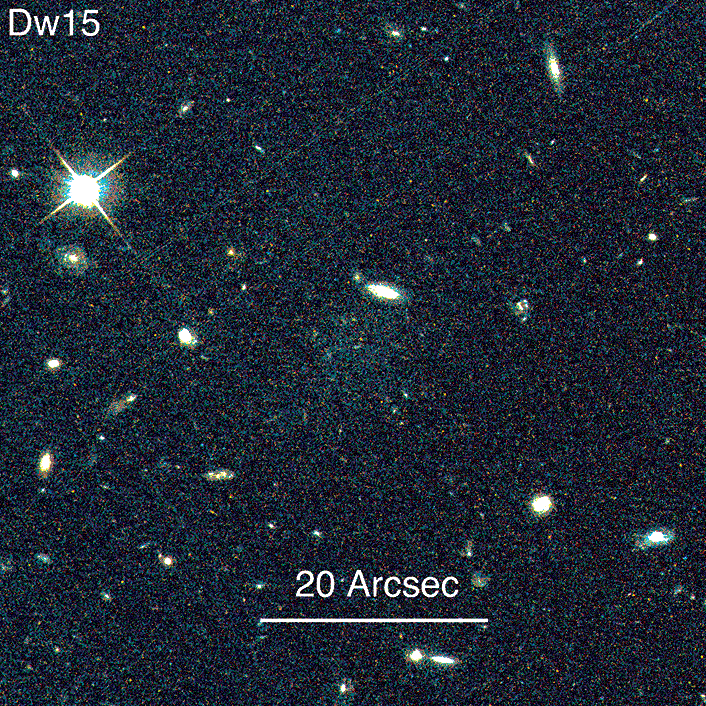}
  \includegraphics[width=7.45cm]{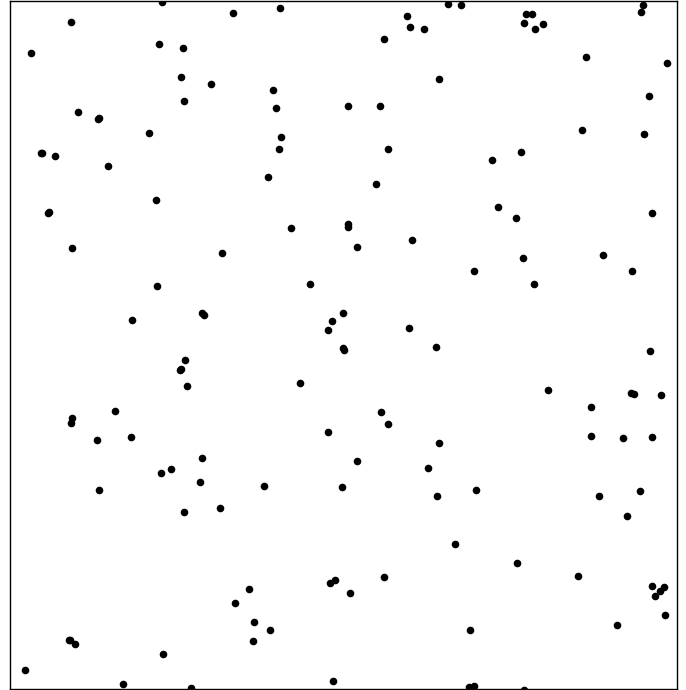}
 \caption{Left: Colorized image cutouts of two unresolved, background dwarfs from {\it HST/ACS}, Dw3 and Dw15.  Right: Plots of all the point sources found by DOLPHOT after quality cuts.  
 Images are 1.0'x1.0', north is up, east is left. There is no apparent overdensity of stars at the position of each dwarf; only DwA and Dw9 show such an overdensity (see Figure~\ref{fig:pic_resolved}). These dwarfs also show no overdensity when only considering sources consistent with being RGB stars.
 The images of Dw3 and Dw15 are representative of the 17 total dwarf candidates that only show diffuse emission in {\it HST} imaging. \label{fig:pic_unresolved}}
 \end{center}
\end{figure*}

The 50\% completeness limit for the {\it HST} observations is at $F814W$=26.9 mag, a value that is consistent across all of the data.  Given a  $M_{I}^{TRGB}$$\approx$$-$4.0 mag \citep[][and references therein]{Gallart05, Radburn11},  we can estimate that any undetected TRGB must have a distance modulus $\gtrsim$30.9 mag and therefore a distance of $\gtrsim$15.1 Mpc for our unresolved dwarf candidate population.
While some of the fainter targets in the unresolved population are hard to see in the {\it HST} imaging, all are visible after spatial binning or smoothing. 
As an illustration of the difference between the resolved and unresolved dwarf samples, we present color images along with point source maps in Figures $\ref{fig:pic_resolved}$ \& $\ref{fig:pic_unresolved}$. 

In addition to the 19 objects we observed with {\it HST} we also examine the {\it HST} data for the unresolved objects from \cite{Merritt16} to update the photometry of these diffuse dwarfs using our techniques. Meanwhile we adopt the \citet{Danieli17} distances and luminosities of M101 DF1, DF2 and DF3 -- the resolved dwarfs from the Dragonfly survey. 

These two populations, resolved and unresolved, will be used to analyze the M101 LF.

To enable comparisons between {\it HST}, CFHTLS \citep{bennet17} and Dragonfly \citep{merritt14} we convert many of our measurements to the $V$ band. 
To convert between the $F606W$ and $V$ band we adopt the method from \cite{Sahu14}  (m$_V$=m$_{F606W}$+0.194), and for the conversion between the $g$ and $V$ band (for the CFHTLS and Dragonfly data) the method from \cite{Jester05} is used:

\textbf{\begin{equation}
m_V=m_g-0.58\times(m_g-m_r)-0.01
\end{equation}}

\subsection{Resolved objects: DwA \& Dw9}\label{subsec:resolved}

Here we discuss the physical and star formation properties of the two resolved dwarfs in our {\it HST} sample, DwA and Dw9; these properties are tabulated in Table~\ref{tab:obj_res}.

\begin{figure*}
 \begin{center}
 \includegraphics[width=6.8cm]{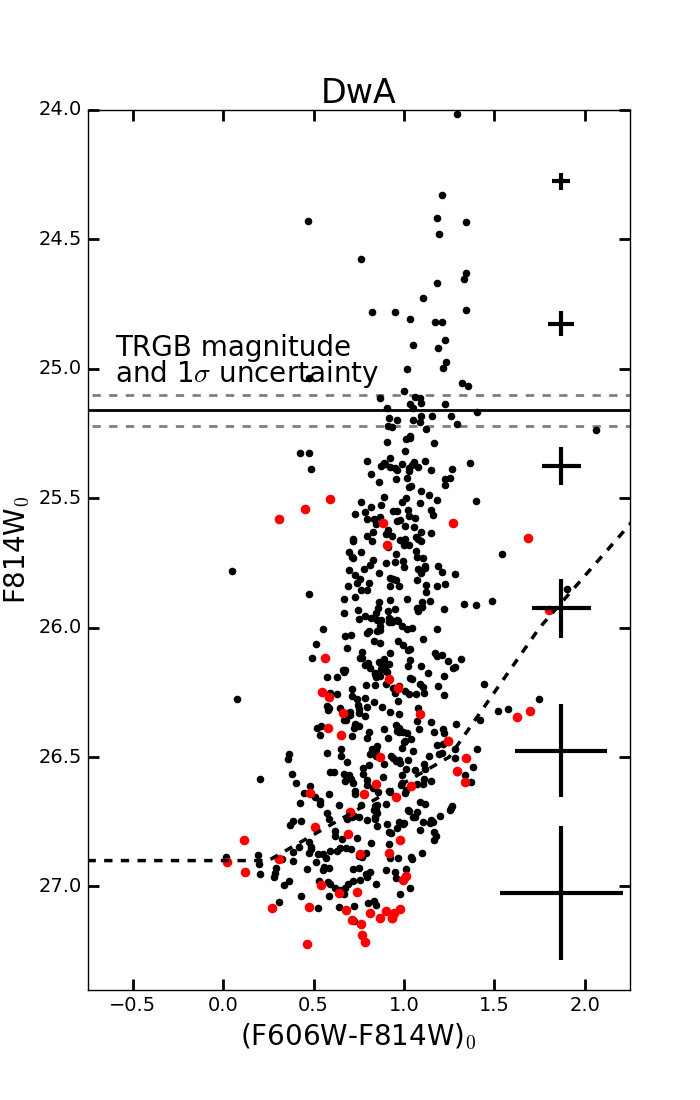}
 \includegraphics[width=6.8cm]{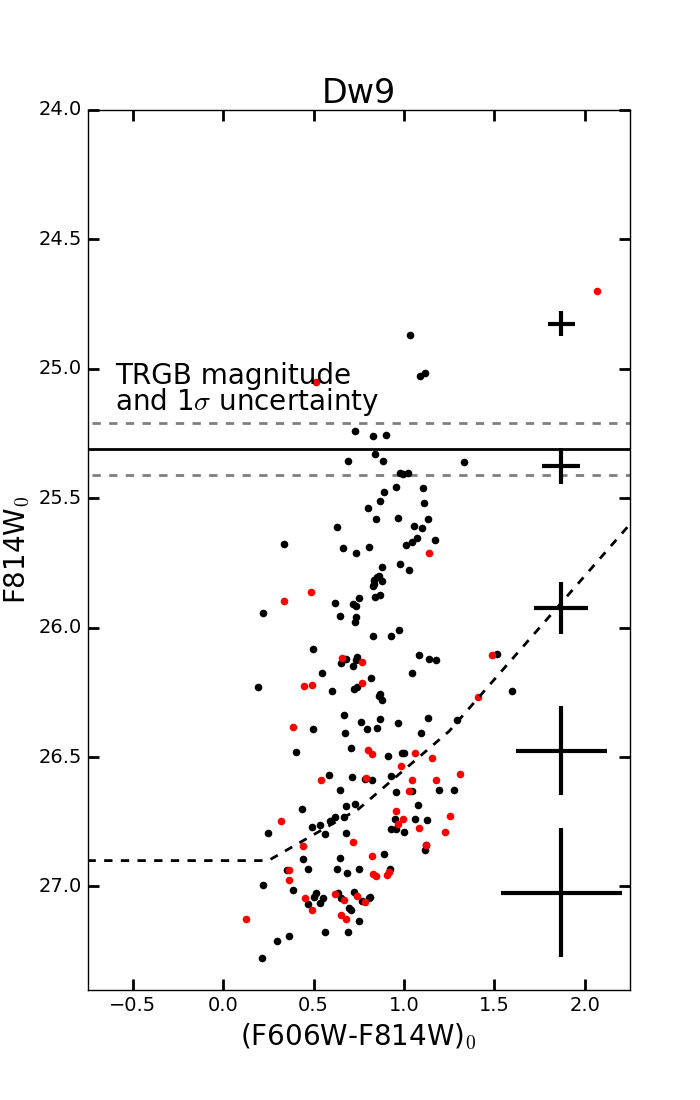}
 \includegraphics[width=6.8cm]{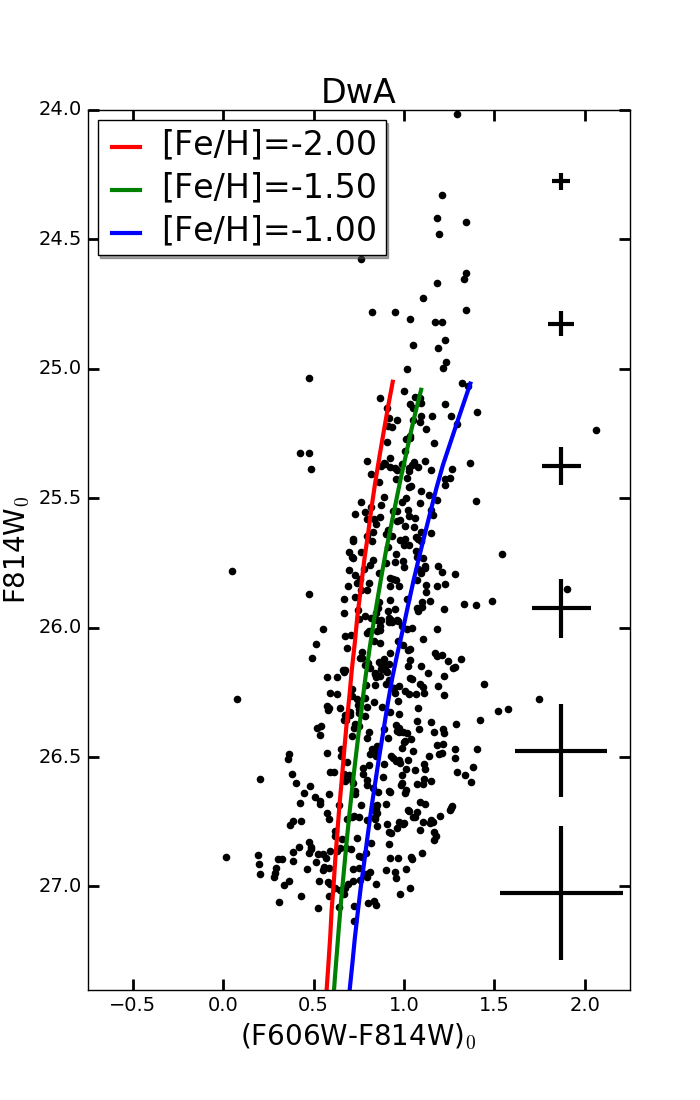}
 \includegraphics[width=6.8cm]{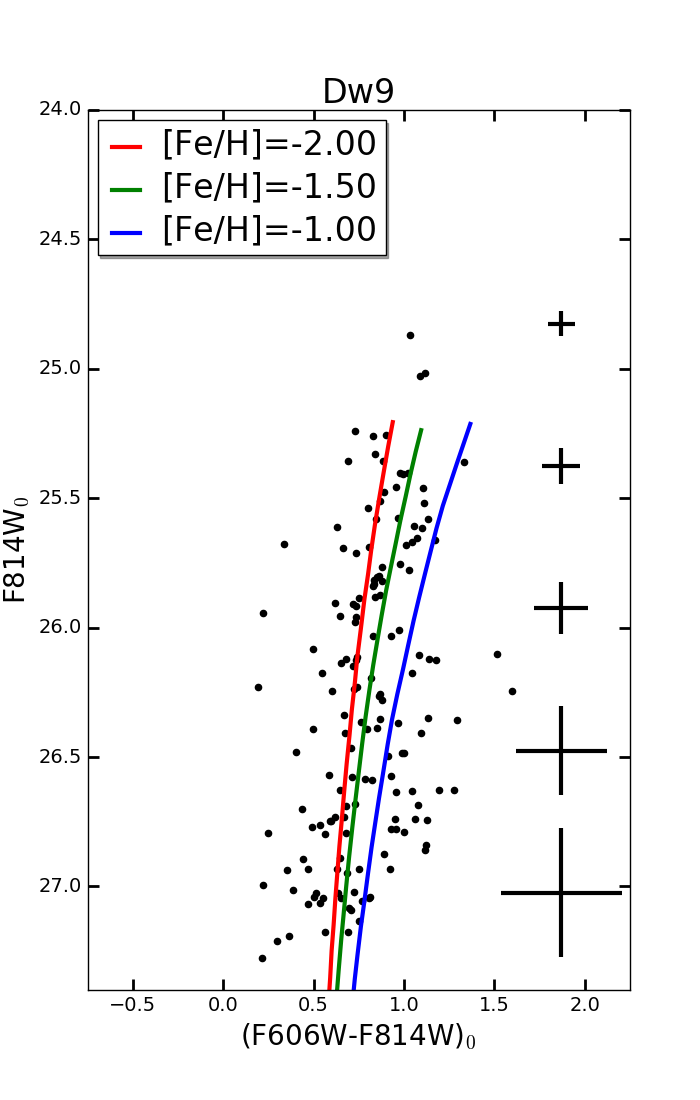}
 \caption{Top: The CMD for the resolved M101 dwarfs, DwA (left) and Dw9 (right). Black dots are point sources centered on each dwarf, while the red dots are stars from an equal area region on the other ACS chip.  The upper black line indicates the TRGB magnitude and 1$\sigma$ uncertainty. The lower dashed line indicates the 50\% completeness limit. Photometric uncertainties are shown along each CMD.  Bottom: The resolved CMDs of DwA (left) and Dw9 (right), with theoretical isochrones at an age of 12.7 Gyr overplotted with varying metallicities \citep{Marigo17}. Old stellar populations with a mean [Fe/H]$\approx$$-$1.5 are consistent with the data. \label{fig:CMD}}
 \end{center}
\end{figure*}

To determine distances to the resolved objects, we make use of the TRGB technique \citep[e.g.,][]{dacosta90,lee93}. The peak luminosity of the RGB is a standard candle in the red bands, because it is driven by core helium ignition and so it provides a useful distance estimate for galaxies with an old stellar component which are close enough that the RGB can be resolved. To determine TRGB magnitudes, we adopt the methodology described in \cite{Crnojevic19}. Briefly, the photometry is first corrected to account for the color dependence of the TRGB \citep{jang17}. Then the field (background+foreground) contamination as derived from a dwarf-free region of the ACS field-of-view is statistically subtracted from the dwarf's CMD. The luminosity function for RGB stars with colors in the range $0.8<(F606W-F814W)_0<1.3$ is computed, and a model luminosity function (convolved with the appropriate photometric uncertainty, bias and incompleteness function derived for the observations) is fit to it with a non-linear least squares method. The uncertainties are derived by re-computing the TRGB for 100 realizations of the statistical decontamination process.
Using this method for the two objects that are resolved into stars in our {\it HST} dataset, DwA and Dw9, we obtain TRGB distances of 6.83$^{+0.27}_{-0.26}$ and 7.34$^{+0.39}_{-0.37}$ Mpc respectively, confirming their association with the M101 group (D $\sim$ 7~Mpc). We show the CMDs for DwA and Dw9 in Figure~\ref{fig:CMD}, along with the TRGB placement for each.

The resolved stellar populations of DwA and Dw9 both appear to be consistent with a stellar population of old, metal poor RGB stars ($\gtrsim$10 Gyr).  In the bottom panels of Figure~\ref{fig:CMD} we plot isochrones \citep{Marigo17} with an age of 12.7 Gyrs. From the CMDs, both dwarfs seem to host stars with mean metallicities of [Fe/H]$\approx$$-$1.5;  the color spread seen in the RGB stars (as derived from a simple Gaussian fit to the RGB sequence) is slightly larger than the photometric uncertainties derived from the ASTs. This could be explained by some intrinsic metallicity spread, however the origin and implications of the spread are beyond the scope of this work. 
DwA has a small population of stars above the TRGB, indicative of asymptotic giant branch stars with intermediate ages ($\sim$2--5 Gyrs based on their magnitudes); such populations are often seen in similarly faint dwarf galaxies in other systems \citep[e.g. Dw2 in Centaurus A;][]{Crnojevic19}.   

Both DwA and Dw9 were not detected in deep 
NUV GALEX imaging \citep{martin05} -- this lack of NUV emission indicates that these galaxies have an upper limit of $\lesssim$1.7$\pm$0.5x10$^{-3}$  M$_{\odot}$/yr for recent star formation, obtained using the relation from \cite{iglesias06}. 
We have performed follow-up HI observations of DwA with the Robert C. Byrd Green Bank Telescope (Karunakaran et al., in prep.) and find no significant HI signal along the line-of-sight to DwA (Dw9 was not observed). We place stringent 5 $\sigma$ upper-limits on its HI mass, log(M$_{HI}$/M$_\odot$)$<$5.75, and its gas-richness, M$_{HI}$/L$_V$=0.68 M$_\odot$/L$_\odot$. The lack of both HI gas and NUV flux is consistent with the old, metal-poor stellar population of DwA. 

The structural properties of the resolved candidates were determined with the maximum-likelihood technique of \citet{Martin08} using the implementation of \citet{Sand12}. The stars selected for the structural analysis are those consistent with the RGB as seen in Figure~\ref{fig:CMD}. We fit a standard exponential profile plus constant background to the data, with the following free parameters: the central position (RA$_0$, DEC$_0$), position angle, ellipticity, half-light radius ($r_h$) and background surface density.  Uncertainties on structural parameters were determined by bootstrap resampling the data 1000 times, from which 68\% confidence limits were calculated.  Key derived parameters are shown in Table~\ref{tab:obj_res}.  Our {\it HST}-derived half-light radii are slightly larger than those found in the ground-based CFHTLS data, likely due to the superior detection of outlying stars at {\it HST} depths and resolution.

The absolute magnitude of the dwarfs is derived via the procedure laid out in \cite{Crnojevic19}. We simulated a well-populated CMD for each dwarf using Padova isochrones \citep{Marigo17} with an age of 12.7 Gyr and metallicity of [Fe/H]=$-$1.5 for each target, assuming a Kroupa IMF \citep{Kroupa01}. This simulated population of stars is then convolved with the photometric errors derived from the artificial star tests.  We then drew stars randomly from this fake stellar population, scaling the number of stars such that it matched the number seen in the RGB region for our observed dwarfs.  The flux from the drawn stars was summed along the entire luminosity function, including stars too faint to be detectable in our $HST$ data, in order to account for the faint unresolved component of each galaxy.  This process was repeated 100 times to assess our uncertainties.  Measurements were converted to the $V$-band using the prescriptions in \citet{Sahu14}.  In Table~\ref{tab:obj_res} we show our derived absolute magnitudes, stellar masses, and we also show our brightness measurements derived from the CFHTLS ground-based data set, as presented in \citet{bennet17}; the data agree to within the errors.

\begin{figure*}
 \begin{center}
 \includegraphics[width=12cm]{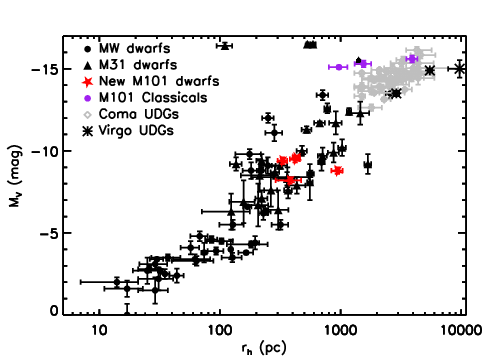}
 \caption{Absolute $V$-band magnitude as a function of half-light radius for M101 dwarf galaxy members as compared to the Local Group and ultra-diffuse galaxies in the Coma and Virgo clusters. The previously known classical M101 group members \citep{tikhonov15} are shown in purple circles and the updated properties for the new M101 dwarf satellites from this work and from \cite{Danieli17} are shown as red stars.  The new population of faint M101 dwarfs is consistent with the size-luminosity relation found in Local Group dwarfs. The data for the MW and M31 dwarf galaxies (black points and triangles, respectively) come from: \citet{mcconnachie12,Sand12,Crnojevic14,drlica15,Kim15a,Kim15b,Koposov15,laevens15a,laevens15b,Martin15,crnojevic16,Drlica16,Torrealba16a,Carlin17,Mutlu18,Koposov18,Torrealba18}. The Coma and Virgo diffuse galaxy properties (gray diamonds and black asterisks, respectively) are from \citet{mihos15,vanDokkum15}.  \label{fig:mv_rh}}
 \end{center}
\end{figure*}

From the properties derived above we can see that these M101 dwarf galaxies, including the dwarfs from \cite{Danieli17}, fit into the Local Group size-luminosity relation, see Figure \ref{fig:mv_rh}.

\subsection{Unresolved objects}

\begin{figure*}
 \begin{center}
 \includegraphics[width=8.7cm]{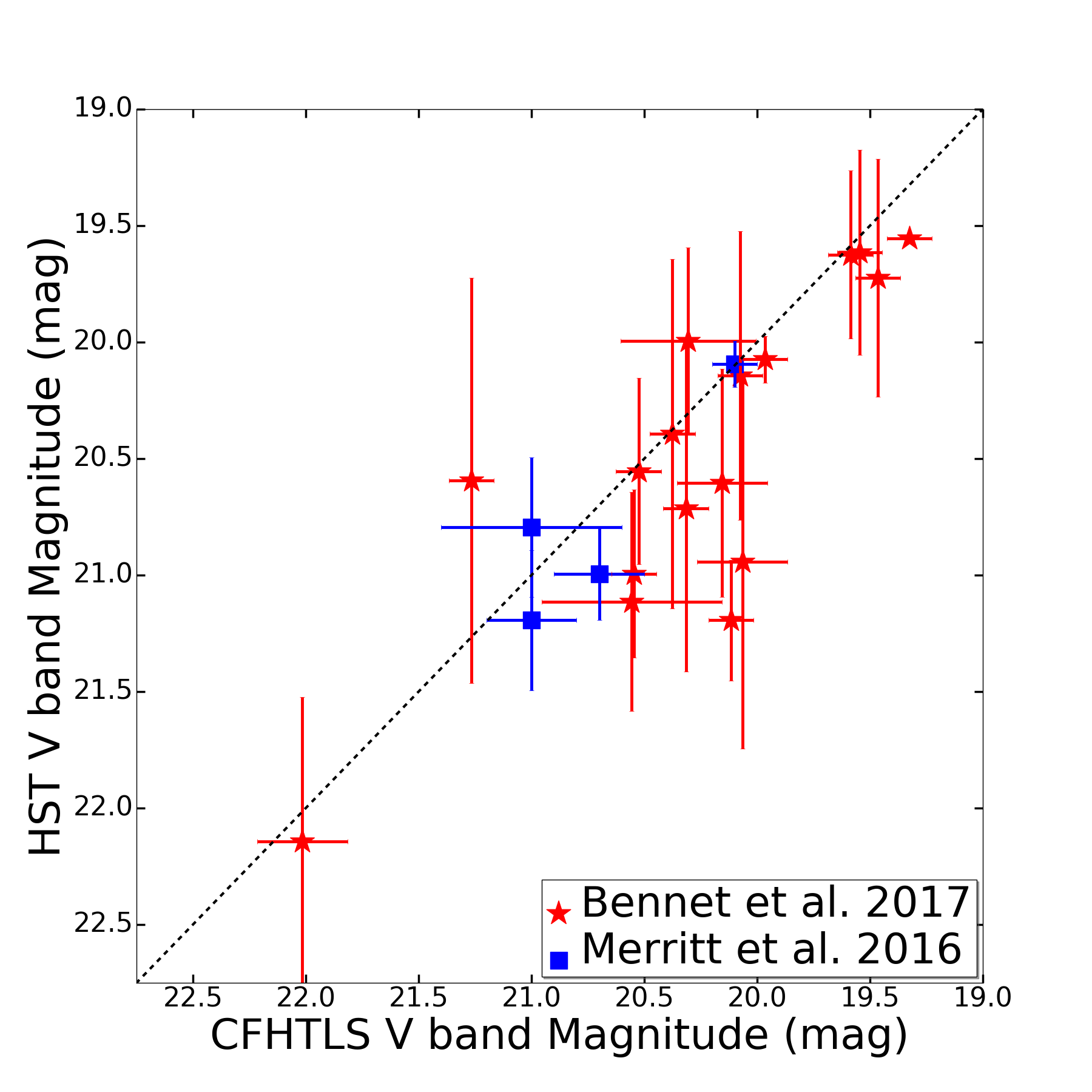}
 \includegraphics[width=8.7cm]{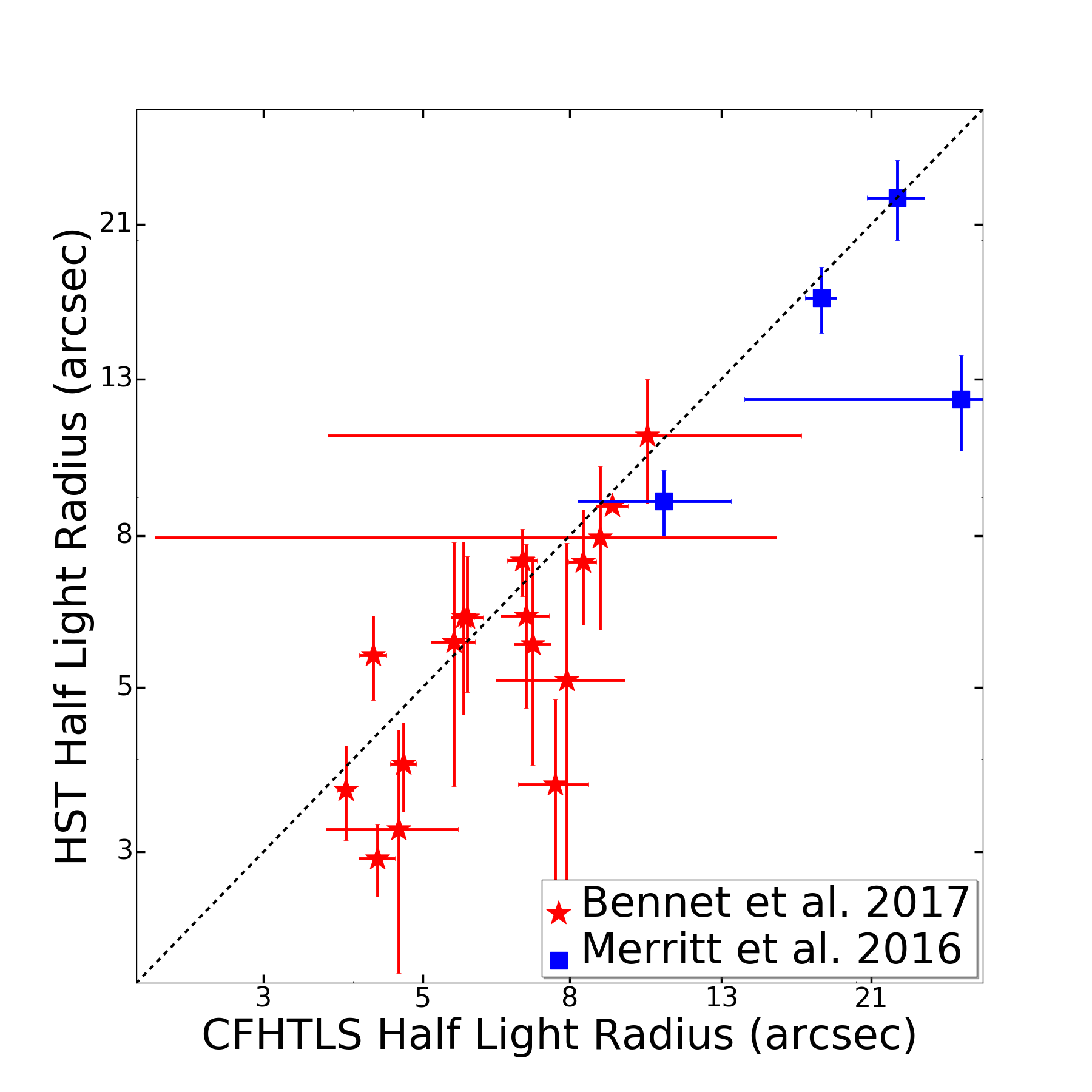}
 \caption{For the 17 diffuse dwarfs imaged with {\it HST}, we compare our {\it HST} derived V-band magnitudes (left) and half light radii (right) with the ground-based CFHTLS measurements made using the techniques of \citet{bennet17}. The measurements made using the two different sets of data are largely consistent. 
 \label{fig:comp}}
 \end{center}
\end{figure*}

The 17 remaining dwarf galaxy candidates imaged with {\it HST} all had unresolved, diffuse emission and are thus in the background and not associated with M101. We remeasured the observational properties of these diffuse dwarfs with the {\it HST} data using GALFIT \citep{peng02}, and the procedure outlined in \citet{bennet17}, including inserting simulated diffuse dwarf galaxies to assess our uncertainties. We also followed this procedure on the unresolved objects from \cite{Merritt16} and found successful fits for all objects aside from DF6, where aperture photometry was used instead. When using GALFIT all parameters were left free.
The {\it HST} images were often spatially binned to increase the signal in each pixel and facilitate the GALFIT measurements; the results are reported in Table \ref{tab:obj_unres}. Comparisons between {\it HST} and the CFHTLS measurements presented in \citet{bennet17} are shown in Figure \ref{fig:comp}, highlighting a good agreement between the values derived from the two datasets.

As can be seen in Figure~\ref{fig:pos}, many of the 17 diffuse {\it HST} dwarfs are projected near the background NGC 5485 group. This is a bimodal group focused around the massive elliptical galaxies NGC 5485 and NGC 5473 at D$\sim$27 Mpc \citep{tully15}. This background group shows a large number of spectroscopically confirmed satellites among the population brighter than our diffuse dwarf sample \citep{makarov11}, and we further constrain its LF in Section~\ref{sec:disc}.

Finally we note that target Dw15, which was regarded as a possible M101 group member after the SBF measurements of \citet{carlsten19}, was found to be unresolved in our {\it HST} data (see bottom row of Figure~\ref{fig:pic_unresolved}).  The SBF measurements successfully show that 14 of the 16 remaining unresolved objects are not M101 group members, with the final 2 (DwB and DwC) being regarded as unlikely M101 group members but not completely excluded. This shows that while the SBF technique reported in \cite{carlsten19a} is a powerful tool, it is still vital to obtain {\it HST} quality follow-up of fainter diffuse dwarf systems to verify their identity and thus refine this promising technique. 

\subsection{Objects not targeted by HST} \label{subsec:obj_not_tar}

Twenty-three dwarf candidates from the \cite{bennet17} sample were not observed by {\it HST} and therefore have unknown distances, aside from Dw26 which was confirmed to be a background galaxy via HI observations \citep{bennet17}. 
The untargeted objects have mean properties (e.g. apparent magnitude and half light radius) that are consistent with the mean properties of the diffuse dwarfs targeted with {\it HST} to within $\sim$1$\sigma$. The observational properties of the untargeted objects as derived from the CFHTLS data are reported in Table 1 of \cite{bennet17}. 

We statistically assess the membership status of the untargeted diffuse dwarfs as a population, based on the {\it HST} observations already made. For this we exclude Dw26 from the sample, as we already know its distance (D$\sim$150 Mpc) due to its HI detection.

If we assume that the 19 objects chosen for {\it HST} follow up are a representative sample of all the LSB objects detected in \cite{bennet17}, then we can determine the number of untargeted objects that are associated with M101 via a series of hypergeometric distributions \citep{Zhang09}. 
Hypergeometric distributions describe a population of size N with a fixed number of successes K (in this case K would be the number of LSB objects that are associated with M101), where k successes are found for every n drawings from the population. 
This is done without replacing previously observed objects back into the population (which would represent a binomial distribution). We use these distributions to determine the likelihood of having 2 'successes' in 19 observations, for various underlying populations with N fixed at 41 (the total population of the \citealt{bennet17} objects once Dw26 is excluded) and K allowed to have any value between 2 and 24 (with a minimum of 2 M101 detections from DwA and Dw9 and a maximum of 24 if all untargeted objects were M101 dwarfs). 
Once the likelihood of each population is determined, we construct a normalized probability function for the values of K. This function is then used to calculate the mean and standard deviation for K.  
These results show that the total LSB population from \cite{bennet17} has 5.1$\pm$2.3 `successes'. As we have already detected two objects associated with M101, this implies that of the remaining 22 untargeted LSB objects 3.1$\pm$2.3 could be associated with the M101 group and 18.9$\pm$2.3 could be unresolved and either associated with the NGC 5845 group or distant background galaxies.  

Another method to estimate the distances of the untargeted LSB objects in \cite{bennet17} is through the surface brightness fluctuations (SBF) method reported in \cite{carlsten19}. These distances show a total of four possible M101 dwarfs (Dw21, Dw22, Dw23 \& Dw35) among the non-targeted LSB objects from \cite{bennet17}, as their reported distances are consistent with the M101 group to within 1 $\sigma$.  
However there are significant uncertainties on these distances and no firm conclusions about the group membership of these objects were drawn by \cite{carlsten19}.  
There are a further 4 objects (Dw19, Dw24, Dw31 \& Dw32) that have reported distances behind the M101 group, but are consistent with M101 group membership within 2$\sigma$. 
From this we can conclude that there are plausibly $\sim$4 M101 dwarfs among the untargeted candidates, with an extreme upper limit of 8 new group members. These numbers are broadly consistent with the results from the statistical method explained above. 

All of these untargeted objects fall on the Local Group size-luminosity relation if they are assumed to be at either M101 (D$\sim$7 Mpc) or NGC 5485 (D$\sim$27 Mpc)  distance. 

The effect that the two different methods used to statistically determine group membership have on the M101 LF is discussed further in Section \ref{subsec:M101_LF}, and can be seen in Figure \ref{fig:LF_NGC}. On the other hand the LF of the NGC 5485 group shows no significant variation between the two methods and therefore this will not be discussed further.

\subsection{M101 Dwarf Galaxy Completeness}\label{subsec:comp}

We have examined the dwarf galaxy completeness around M101 on both the bright and faint ends. For a bright limit on the completeness of the \cite{bennet17} sample, we tested the detection algorithm used in that work via artificial dwarf galaxy simulations, to an apparent magnitude of m$_g$=17 (M$_g$=$-$12.2 or $M_V$=$-$12.5 at 7 Mpc) on the bright end.

At first blush, the bright limit of \cite{bennet17} suggests it is possible that bright satellites of M101  ($M_V$$<$$-$12.5) may still be undiscovered, but this is unlikely: previous studies around M101 \citep[e.g.][]{karachentsev13} were considered complete down to M$_B$=$-$11.2 ($M_V$=$-$12.1), which overlaps with the Bennet et al. bright end brightness limit, and is explicitly complete to $M_V$$\approx$$-$20 and possibly brighter \citep[see e.g. Figure 5 of ][]{karachentsev13}.  
We are exploring additional modifications to our algorithm to probe an expanded parameter space (Bennet et al. in prep.): to date, preliminary results around M101 indicate no additional dwarf candidates up to a brightness of m$_g$=16 (M$_g$=$-$13.2 or $M_V$=$-$13.5).

For our faint end completeness limit to M101 dwarfs, we note that the 50\% completeness limits of the \citet{bennet17} diffuse dwarf search was $M_g$=$-$7.2, or $M_V$=$-$7.5.  Several of the faintest dwarf candidates in that work were not targeted with {\it HST}, but have SBF distances that are at least consistent with M101 (Dw21, Dw22, Dw23 \& Dw35); these dwarfs should be followed up with future space-based imaging.  {\it HST} imaging is available for all diffuse dwarfs with SBF distance estimates consistent with M101 \citep{carlsten19} and with magnitudes $M_V\lesssim-$8.2 mag (DwA, Dw9 \& Dw15): we thus adopt this value as a conservative faint limit for our completeness.

\section{Discussion} \label{sec:disc}

Understanding the behavior of the galaxy satellite LF towards the faint end is crucial in constraining the formation and evolution of galaxies. It is also important in furthering our understanding of the relation between the stellar and dark matter content in dwarf galaxies. 

The observed LF slope around the MW ($\alpha \sim$$-$1.2 assuming a Schechter function,  e.g., \citealt{Koposov08}) is far shallower than what is predicted based on the mass function of dark matter subhalos  from simulations ($\sim$$-$1.9, e.g., \citealt{Bullock17}). Several possible explanations, including observational incompleteness and/or theoretical modeling have been proposed in the past decade to address this issue \citep[e.g.][]{Tollerud08,Brooks13,Hargis14,Sawala16,Wetzel16,Garrison17,Kim17} -- the general consensus is that the incorporation of mechanisms such as feedback, star formation efficiency and re-ionization into cosmological simulations can help reconcile the differences between the observed slope of the LF and  theoretical predictions. 
However, such models often aim to reproduce the singular case of the MW LF, which may not be representative. Thus, in order to test the robustness of these models against a larger sample, it is necessary to observe the faint end of the satellite LF of systems beyond the Local Group. Such observations will allow us to probe the typical value of the LF slope at lower luminosities, as well as constrain the system--to--system scatter.

\subsection{The Satellite Luminosity Function of M101}\label{subsec:M101_LF}

The satellite LF for M101 was constructed by using all galaxies that are reported to be within the projected virial radius of M101 ($\sim$250 kpc or $\sim$2.05 degrees at 7 Mpc) and with confirmed distances consistent with that of M101. 
For this reason we do not include the proposed M101 members DDO194 ($D_{proj}$=656 kpc, \citealt{tikhonov15}) or KKH87 ($D_{proj}$=392 kpc, \citealt{karachentsev13}) as these objects are projected too far from M101. We also do not include UGC08882, as the reported distance (D=8.3$\pm$0.8 Mpc, \citealt{Rekola05}) places the object behind M101. We do include the three faint galaxies (DF1, DF2 \& DF3) found by the Dragonfly survey that were confirmed via TRGB distance measurements (\citealt{merritt14}, \citealt{Danieli17}).  See Table \ref{tab:M101} for a full list of M101 members that we consider, including those confirmed by the current work.

\floattable
\begin{deluxetable}{c|cc|c|cc|c}
\tablecaption{Confirmed M101 group members within 250 kpc\label{tab:M101}}
\tablehead{
\colhead{Name} & \colhead{RA} & \colhead{Dec} & \colhead{M$_V$} & \colhead{Projected M101} & \colhead{Distance} & \colhead{Ongoing Star}\\
\colhead{} & \colhead{} & \colhead{} & \colhead{} & \colhead{Distance (kpc)} & \colhead{(Mpc)} & \colhead{Formation} }
\colnumbers
\startdata
M101 & 14:03:12.5 & +54:20:56 & -20.8 & 0 & 6.79$\pm$0.41\tablenotemark{a} & Y \\
NGC 5474 & 14:05:01.6 & +53:39:44 & -17.6 & 89 & 6.82$\pm$0.41\tablenotemark{a} & Y \\
NGC 5477 & 14:05:33.3 & +54:27:40 & -15.3 & 44 & 6.77$\pm$0.40\tablenotemark{a} & Y\\
Holm IV & 13:54:45.7 & +53:54:03 & -15.0 & 160 & 6.93$\pm$0.48\tablenotemark{a} & Y\\
M101 DF1 & 14:03:45.0 & +53:56:40 & -9.6 & 50 & 6.37$\pm$0.35\tablenotemark{b} & N \\
M101 DF2 & 14:08:37.5 & +54:19:31 & -9.4 & 97 & 6.87$^{+0.21}_{-0.30}$\tablenotemark{b} & N \\
M101 DF3 & 14:03:05.7 & +53:36:56 & -8.8 & 89 & 6.52$^{+0.25}_{-0.27}$\tablenotemark{b} & N\\
M101 DwA & 14:06:49.9 & +53:44:30 & -9.5 & 100 & 6.83$^{+0.27}_{-0.26}$\tablenotemark{c} & N \\
M101 Dw9 & 13:55:44.8 & +55:08:46 & -8.2 & 160 & 7.34$^{+0.39}_{-0.37}$\tablenotemark{c} & N \\
\enddata
\tablenotetext{a}{From \cite{tikhonov15}}
\tablenotetext{b}{From \cite{Danieli17}}
\tablenotetext{c}{From this work.}
\end{deluxetable}

In order to place the M101 LF into context, we also compile the cumulative LFs of other  Local Volume galaxies, both inside the Local Group (M31 and the MW itself) and outside of it (M94, M81 and Centaurus A), using data from recent resolved stellar population studies. 
All of the galaxy groups have been examined for satellites down to at least $M_V$=$-$9.1, which allows for the widest possible comparison down to the limiting magnitude reached in the M101 survey. 
These hosts span a range of masses from 2.5$\times$10$^{11}$$M_{\odot}$ for the MW \citep{Eadie16} to 9.0$\times$10$^{11}$$M_{\odot}$ for Cen A \citep{Woodley10} using dynamic mass estimates for the innermost 40 kpc (where these estimates are available). 
From this sample there is no evident correlation between host mass and satellite LF, which is likely due to the relatively small mass range examined. These galaxies also span a range of environments, from very isolated groups like M94 to groups that are actively accreting smaller galaxy groups, like suggested for the MW and the LMC and possibly for
M81 accreting M82 and NGC3077 \citep{chiboucas13}. The galactic environment seems to be more important in predicting the galaxy LF as will be discussed in Section \ref{subsec:density} in more detail. 

For the compilation of these LFs, we have only used objects where the association between the dwarf galaxy and the host has been confirmed via distance measurements. We do not include dwarfs fainter than $M_g$=$-$7.2 ($M_V$=$-$7.5), which corresponds to the 50\% completeness limit for the survey of M101 reported in \cite{bennet17}. Note that we consider our M101 sample complete down to $M_V$=$-$8.2, as discussed in Section~\ref{subsec:comp}, but for the statistical corrections for those dwarfs not imaged with {\it HST} we use the full \cite{bennet17} sample down to $M_V$=$-$7.5. We also limit our analysis to satellite galaxies within a projected distance of $<$250 kpc from the host, or with 3D distance of D$<$250 kpc in the case of the MW.

We consider several sources for the satellite galaxies used to construct our cumulative LFs, following \cite{Crnojevic19}. Briefly, for the MW we adopt the updated online 2015 version of \cite{mcconnachie12} -- while this excludes some recent ultra-faint discoveries, they are below the limiting magnitude of the M101 survey and are thus not considered further. For M31 we combine the catalogues from \cite{martin16} and \cite{McConnachie18}. For hosts beyond the Local Group, we adopt the Updated Nearby Galaxy Catalogue \citep{karachentsev13} and the Extragalactic Distance Database\footnote{http://edd.ifa.hawaii.edu/} \citep{Jacobs09}, complementing these resources with more recent work where available. More specifically, for M81 we refer to \cite{chiboucas13}, who performed a wide field survey of M81 and its satellites with CFHT/Megacam along with {\it HST} follow-up imaging (we excluded objects that are considered tidal dwarfs). We also include the M81 dwarf D1005+68 \citep[M$_V$=$-$7.94;][]{Smercina17}. For M94 we include the new discoveries from \cite{Smercina18}, which adds two faint dwarf galaxies to the two already known M94 satellites, there are also 12 additional objects with velocities consistent with M94 \citep{karachentsev13}, however these objects are outside of our 250 kpc limit.  
For Centaurus A we use the results from \cite{Crnojevic19}, who compiles a Centaurus A LF based on the discovery of 11 new satellites within the PISCeS survey as well as on 13 previously known dwarfs. We note that Centaurus A is an elliptical galaxy, and it likely has a very different accretion history to that of the MW and the other spiral galaxies that we examine here; nevertheless, its mass is considered to be comparable to that of the MW within a factor of a few \citep[e.g.,][]{kara07}. 
Though we reiterate that the dynamic mass estimates for these galaxies (where available) show variation of $\sim$3.5 between the smallest and largest galaxies, this does not appear to correlate with satellite number in the examined galaxy sample.  

The completeness of the different surveys we consider is not easy to quantify.  
Despite efforts to ensure a uniform sample, a few caveats are inevitable: i) MW surveys suffer from incompleteness effects, mainly due to incomplete spatial coverage, especially in the direction of the Galactic plane. Such effects could lead to an underestimate of the number of faint satellites by a factor of $\sim$3 \citep[e.g.][]{Hargis14}, meaning our constructed LF for the MW is a lower limit; ii) \citet{Crnojevic19} estimate the incompleteness of the PISCeS survey and suggest that there might be $\sim$5 to 10 yet undetected galaxies in the range $-10<M_V<-8$ within $\sim150$~kpc from Cen A. This should, however, not significantly alter the slope of the derived LF. At larger galactocentric distances (between 150 and 300~kpc), 13 candidates with $-12<M_V<-8$ have additionally been presented by \citet{Muller17a}, but they have not been confirmed as Centaurus~A members yet; iii) a deep search for faint satellites has not been performed beyond the innermost 150 kpc for M94, thus the LF constructed for this host is also likely a lower limit.
A search within 150 kpc of projected distance includes over $\sim$80\% of the virial volume for a galaxy the size of M94 (considering the 150 kpc projected radius cone that was observed, which will have sensitivity to some satellites at larger 3D radii), and the satellites themselves may be centrally concentrated and distributed like a NFW profile. Nonetheless, we do conduct a complementary analysis by comparing our Local Volume galaxies using only the innermost 150 kpc of projection in addition to our searches across the entire virial radius. 

\begin{figure*}
 \begin{center}
 \includegraphics[width=16cm]{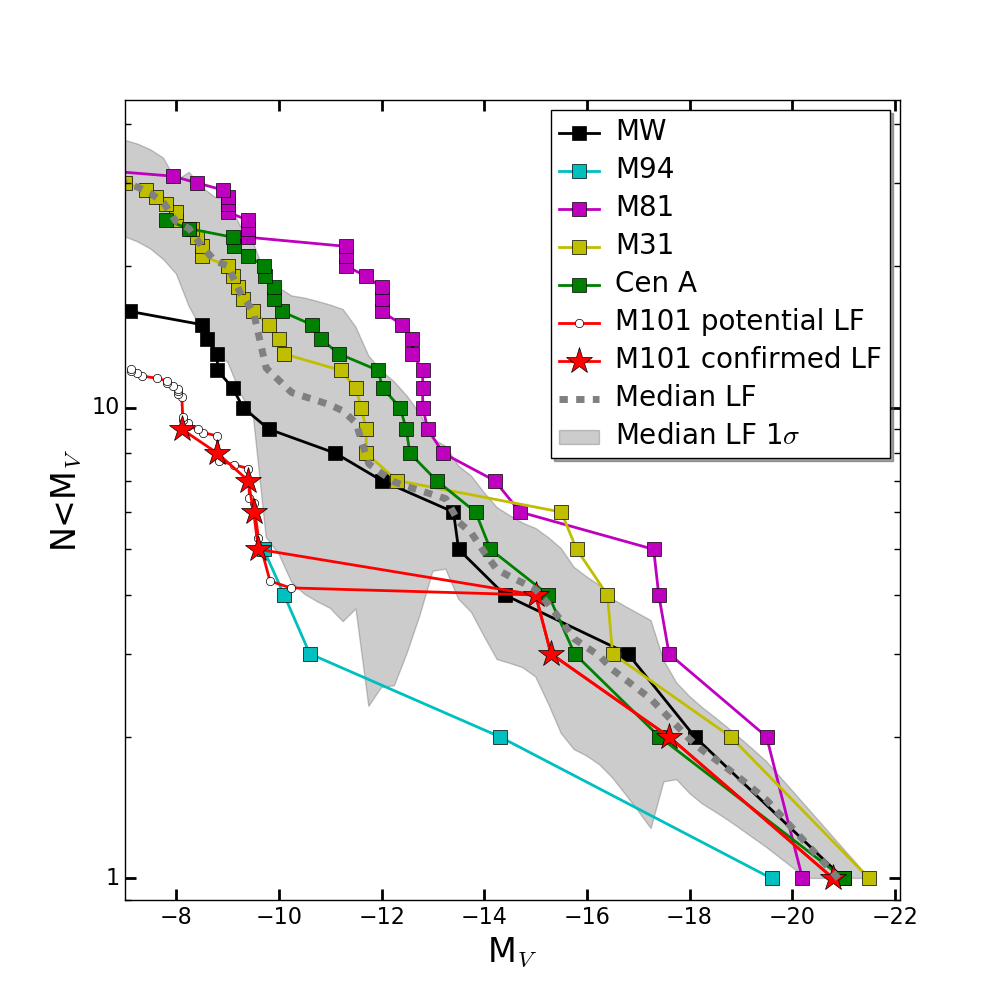}
 \caption{The cumulative satellite LF for several Milky Way-like systems out to a projected radius of 250 kpc, and the constructed median from the set.  The M101 LF is displayed with star symbols, while the small circles indicate the LF for M101 after we include the statistical weighting of targets not followed-up with {\it HST} (see discussion in \ref{subsec:obj_not_tar}).  
 The grey dashed line is the median of the individual galaxy measurements while the shaded region represents the 1$\sigma$ scatter of the median. No attempt was made to correct any LF for incompleteness; we consider the M101 LF complete down to $M_V$$\approx$$-$-8.2 mag for confirmed M101 dwarfs and $M_V$$\approx$$-$-7.5 mag for potential M101 dwarfs. The data for the group LFs come from \cite{Smercina18} for M94 represented by the cyan squares, \cite{Crnojevic19} for Cen A represented by the green squares, 
 \cite{chiboucas13} and \cite{Smercina17} for M81 represented by the magenta squares, \cite{martin16} \&  \cite{McConnachie18} for M31 represented by the yellow squares and \cite{mcconnachie12} for the MW represented by the black squares. Note that this is a lower limit for the MW due to incomplete spatial coverage. \label{fig:LF_NGC}}
 \end{center}
\end{figure*}

We perform a fit adopting a Schechter function for each cumulative LF using a maximum likelihood estimator:

\begin{equation}
N(<M) = \phi_\ast \gamma [\alpha+1, 10^{0.4(M_\ast-M)}]
\end{equation}

\noindent where $\phi_\ast$ is the normalization density, $\gamma$ represents the incomplete gamma function and depends on the slope ($\alpha$) of the LF and on M$_\ast$, where M$_\ast$ is the characteristic magnitude. 
We find the best fit values for $\alpha$ to be $-$1.25$\pm$0.05 for Cen A, $-$1.19$\pm$0.06 for the MW, $-$1.28$\pm$0.06 for M31, $-$1.21$\pm$0.05 for M81, $\sim$$-$1.18 for M94 (which was difficult to constrain, given how few satellites are in this system) and finally $-$1.14$\pm$0.10 for M101 itself. These slopes are consistent with previous literature results and with each other. 
We also create the LF of the 'median' Local Volume host (see Figure \ref{fig:LF_NGC}): we derive the median number of satellite members at each 0.25 mag increment in the range $-$20.25$\leq$M$_V$$\leq$$-$7.0, these are then assembled into a 'median' LF. 
We then fit this LF using the same method that was used for the individual galaxies, obtaining a slope of $\alpha$=$-$1.29$\pm$0.05. 
While the values of $\phi_\ast$ and M$_\ast$ are not well constrained, variations of these parameters within reasonable limits do not significantly affect the best fit values for $\alpha$ (see \citealt{Chiboucas09} and \citealt{Park17}).

The above values for $\alpha$ were constructed using only confirmed M101 members, and we explore the effects that the unconfirmed diffuse dwarfs would have on this slope. If we include the statistically weighted objects that do not have follow-up imaging (see \ref{subsec:obj_not_tar} for details of the statistical weighting), this adds a total of $\sim 3$ dwarfs over the magnitude range $-7.0\leq$M$_V\leq-$10.2.  
The best fit obtained for this LF is $\alpha=-1.16\pm0.12$. This potential M101 LF can be seen in Figure \ref{fig:LF_NGC}. 
However, if instead of the statistically weighted objects we add those objects that have SBF distances consistent with M101 membership to within 1$\sigma$ (i.e., Dw21, Dw22, Dw23 \& Dw35) to the M101 LF, we obtain a slope of $\alpha=-1.15\pm0.10$.  
Thus it can be seen that the inclusion of the unconfirmed M101 members only gives rise to minor LF changes. 

Figure \ref{fig:LF_NGC} presents a direct comparison of the LFs for the considered Local Volume galaxies. At brighter magnitudes (M$_V$$<-$12), M101 appears similar to the other Local Volume galaxies: specifically, M101 has 4 members with M$_V<-12$, compared to a median value of 7.5$\pm$4.7 for the other groups. 
However at fainter magnitudes these LFs show substantial variation, with each of M81, Cen~A and M31 having $\gtrsim$20 satellites down to M$_V$=$-$8.0, as compared to the 9 satellite members of M101 down to the same magnitude. This uniqueness is borne out when comparing M101 to the median Local Volume galaxy: at M$_V$=$-$8.0, M101 has 9 members compared to 24.5$\pm$7.7 for the median. 
While there are large uncertainties due to small number statistics,  it is still apparent that the M101 system is unusual and sparse. 
This sparseness within the M101 and M94 groups is mirrored by examinations of the innermost 150 kpc only, this shows M101 with 7 members compared to a median of 17.3$\pm$4.1. 

The SAGA survey \citep{Geha17} reported on satellite galaxies around eight MW analogues with distances in the range of 20-40 Mpc, with a limiting magnitude M$_r$$<$$-$12.3. 
These results produced LFs that were comparable to the MW and M31. When we extend this comparison to the galaxies in our sample we find that that the majority of the SAGA galaxies have poorer satellite systems than the median Local Volume galaxy LF. At the SAGA survey's limiting magnitude (M$_V$$<$$-$12.5) the Local Volume galaxies have a median of 6.8$\pm$3.8 group members, this is more group members than all but one SAGA galaxy (NGC 6181) and over half of the SAGA sample are below the 1$\sigma$ limit. 
Therefore we conclude that the LFs found by the SAGA survey more closely resemble those of the M101 and M94 groups than the MW or M31.  

Our data additionally shows a lack of any M101 group members with an absolute magnitude between M$_V$=$-$15 to M$_V$=$-$10. This area of parameter space has been fully probed by previous work in the M101 system out to the projected virial radius \citep[e.g.][]{karachentsev13,karachentsev15,merritt14,bennet17}, and therefore this gap appears to be genuine -- see Section~\ref{subsec:comp} for a discussion of our bright and faint satellite magnitude limits.  
The addition of the 22 objects not targeted by HST does not close this gap (if any of them actually belong to the M101 system), but could potentially decrease the size, as the brightest of these objects (Dw31 \& Dw32) would be brighter than DF-1 (M$_V$=$-$9.6) when placed at the distance of the M101 group. 
However, this still results in a gap of 4.9 mag, far larger than that observed in other comparable nearby galaxies and the SAGA survey.  
The mean largest magnitude gap between any two confirmed group members in this latter sample is 2.6$\pm$1.1~mag, which is far below the 5.4~mag gap recorded for the M101 system (see Figure \ref{fig:gap}). 
A full examination of the theoretical implications of such a large gap is beyond the scope of this work \citep[although see][]{Ostriker19}. However, it should be noted that this behaviour is not observed in galaxy simulations, which tend to produce a satellite LF close to those observed in the MW and M31, with far smaller magnitude gaps \citep{Geha17}. 

\subsection{Satellite Population and Galactic Environment}\label{subsec:density}

Previous work on the environmental dependence of galaxy LFs \citep[for a summary see][ and the references therein]{ferrarese16} has examined a variety of different cluster and group environments. There have been multiple contrasting results with some studies finding consistent slopes across different mass ranges, morphologies and environments, and others finding a density dependence on the faint-end slope of the LFs. 
We find that the derived LF slopes for our galaxies are consistent with each other within the uncertainties, and therefore in this respect do not depend upon galactic environment. 

However, while the derived LF slopes are similar across all tested galaxies,  M101 and M94 clearly have fewer satellite galaxies, particularly at the faint-end, suggesting that the LF slope may not be the best way to compare. 
This sparseness might be the result of the galactic environment as M101 and M94 are in lower density regions, as we will now examine.

The relationship between environment and dwarf galaxy number can be quantified via the tidal index parameter (i.e. density contrast,  \citealt{karachentsev13}): 

\begin{equation}
\Theta_{5} = \log_{10}(\sum_{n=1}^{5} M_{n}/D_{in}^{3})+C
\end{equation}

\begin{figure}
\begin{center}
 \includegraphics[width=8.7cm]{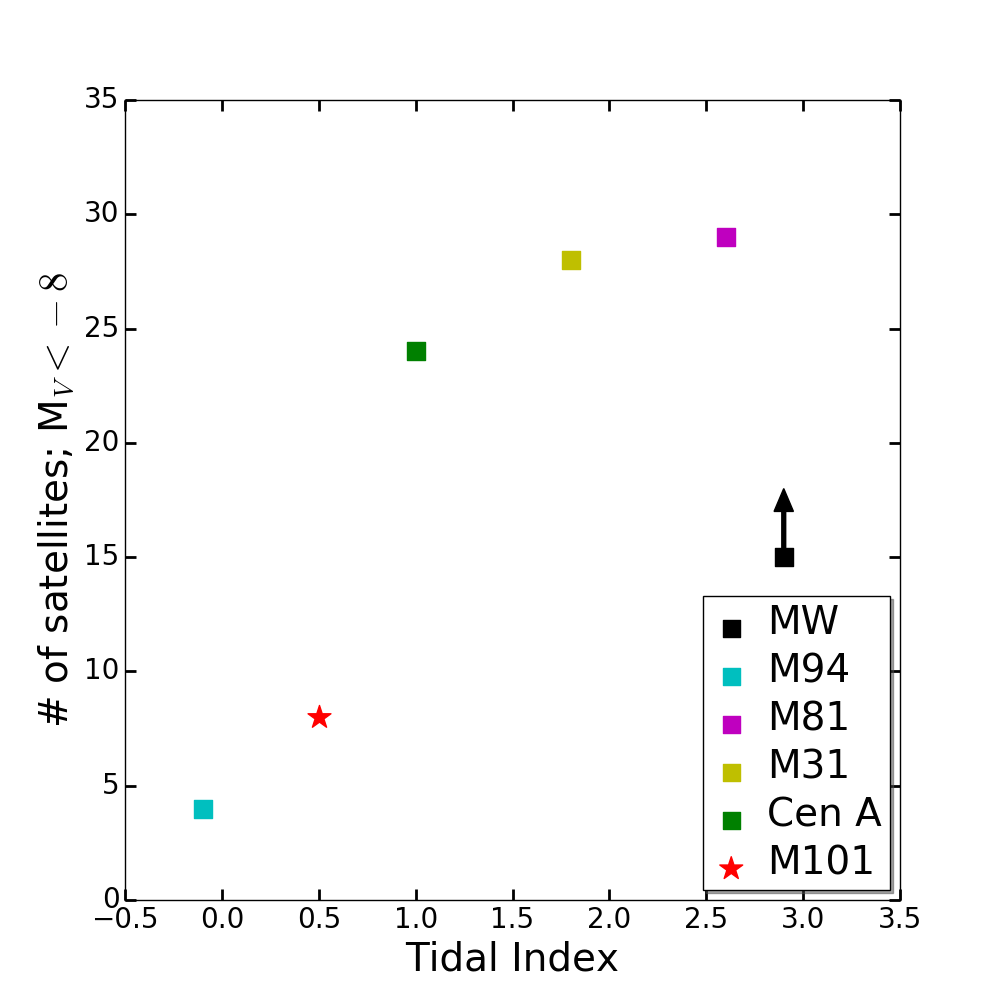}
 \caption{The environmental density of our target galaxies, based on tidal index \citep[i.e. density contrast, where smaller numbers indicate a more isolated galaxy; see][]{karachentsev13}, against number of confirmed satellites with M$_V$$\leq$$-$8. M101 is represented by a red star, the other Local Volume galaxies are represented by squares. The MW (black square) is a lower limit due to the spatial incompleteness of surveys caused by the Galactic plane.
 \label{fig:density}}
 \end{center}
\end{figure}

\noindent This tidal index ($\Theta_{5}$) is calculated using the tidal force magnitude on a galaxy `i' by the neighbouring galaxies `n'. This tidal force magnitude depends on the mass of galaxy $n$, $M_n$, and the 3D separation between the galaxies $D_{in}$. 
C is a constant equal to $-$10.96 which has been chosen such that if $\Theta_{5}$$\leq$0 then the galaxy is isolated. 
The tidal index is the summation of the tidal force magnitude from the five neighbours of a galaxy where this magnitude is the highest. 
We find that M101 and M94 have lower tidal indices (0.5 and -0.1 respectively) than the other Local Volume galaxies examined which are all $\geq$1.0 (see Figure \ref{fig:density}). For this work we draw tidal indices from \cite{karachentsev13} where possible. 
A possible problem with the tidal index is that a satellite galaxy is more likely to be found at the apocenter of its orbit rather than the pericenter; causing the tidal index to be underestimated. However the use of the 5 most influential galaxies should lessen the potential impact of any individual galaxy's orbital positioning. 

In Figure~\ref{fig:density} we plot the number of satellites with M$_V$$\leq$$-$8 as a function of the tidal index of the central galaxy, which appears to show a relationship between tidal index and number of satellites, where the objects with larger tidal indices also have more satellites.  The exception to this proposed relation is the MW which has the largest tidal index of the galaxies examined (2.9), however it has fewer reported satellites than M31, M81 or Centaurus A. This could be explained by the spatial incompleteness in surveys of the MW caused by the Galactic plane. 
We have also examined a more limited sample using only satellites within 150 kpc projected distance, this showed the same trend however with a larger scatter. 
This potential relationship between satellite number and density deserves further attention in future work. 

\subsection{Star formation in Satellite Galaxies}\label{subsec:star_form}

Careful examination of the CMDs for DwA and Dw9 yield no evidence for star formation and show objects that can be fit by a single old population of stars (see Figure \ref{fig:CMD} and discussion in Section~\ref{subsec:resolved}). This is supported by a lack of NUV emission or HI gas associated with either DwA or Dw9, and is consistent with the Local Group where galaxies less massive than the Magellanic Clouds within the virial radius of the MW and M31 are not star forming and have no HI gas \citep{spekkens14}.  

The SAGA survey \citep{Geha17} reported star formation in most (26 of 27) of the satellite galaxies they found down to magnitude M$_r$$<$$-$12.3 around eight Milky Way analogues, in contrast to that observed around M31 and the MW. 
These dwarfs were identified as star forming using the presence of H${\alpha}$. 
Here, we measure the star forming fraction of the satellite galaxies of Local Volume galaxies (M81, CenA, M94 and M101). We have determined which dwarfs are star forming via the presence of bright, blue main sequence stars in resolved imaging.  
At magnitudes brighter than the SAGA magnitude limit (M$_V\lesssim-$12 mag), we find that the Cen A and M81 groups have star forming dwarf fractions that are 30\% (3 of 10, \citealt{Crnojevic19}) and 38\% (5 of 13, \citealt{chiboucas13}), respectively. 
This is comparable to those of the MW (40\%, 2 of 5) and M31 (20\%, 2 of 10). 
On the other hand, we find that M94 (1 of 1, \citealt{Smercina18}) and M101 (3 of 3) both show star formation in all satellites above the SAGA magnitude limit. At fainter satellite magnitudes than that probed by the SAGA survey (e.g. M$_V$$<$$-$12), we observe no star forming dwarfs around our target host galaxies.  The one exception is M94, where we find that all recorded satellites (4 of 4) are star forming \citep{Smercina18} with the faintest at M$_V$=$-$9.7.

This result points to a relationship between environment and star formation fraction among the brightest satellites of the examined Local Volume galaxies. The SAGA galaxies were selected to be isolated, with no galaxies within one magnitude and one degree in projection, and no massive galaxies (5$\times$10$^{12}$M$_\odot$) within 2 virial radii \citep{Geha17}. These isolation requirements may have pushed their sample towards galaxies in low density environments more akin to M101 or M94 than the MW. 

\begin{figure*}
\begin{center}
 \includegraphics[width=8.7cm]{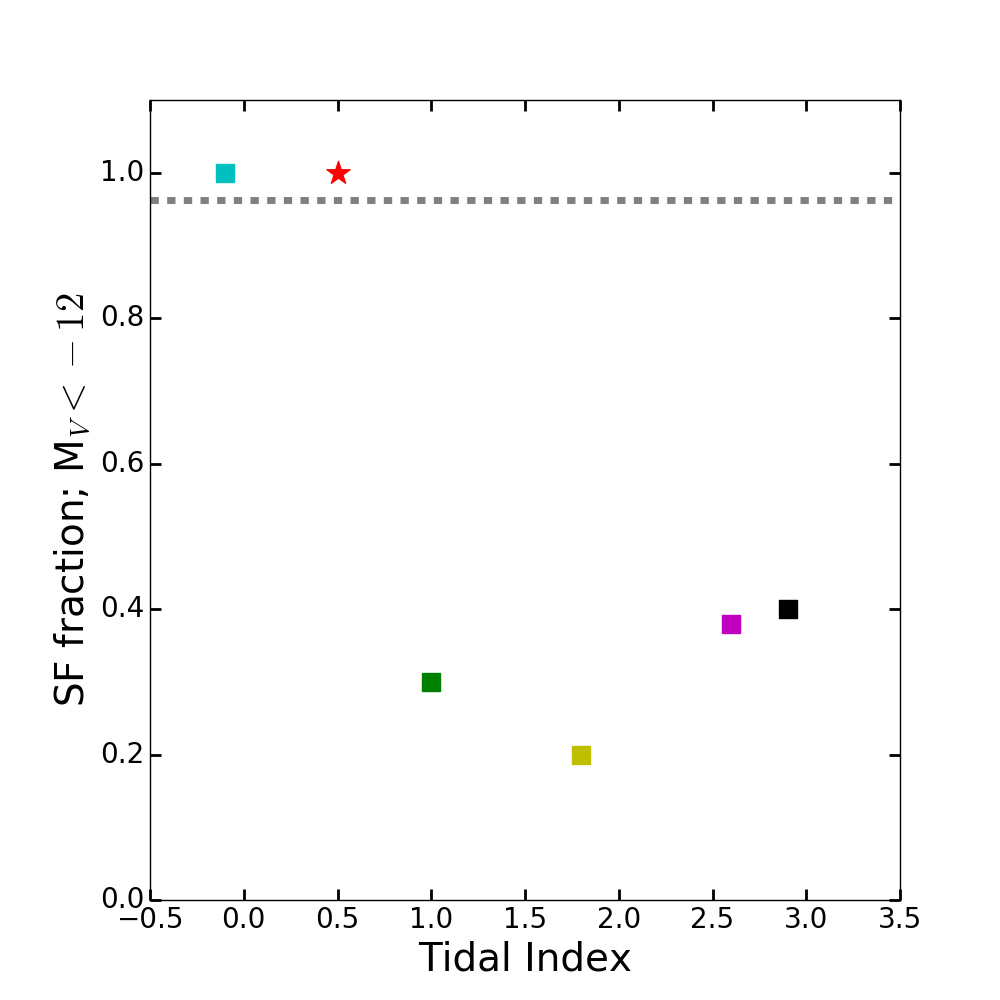}
 \includegraphics[width=8.7cm]{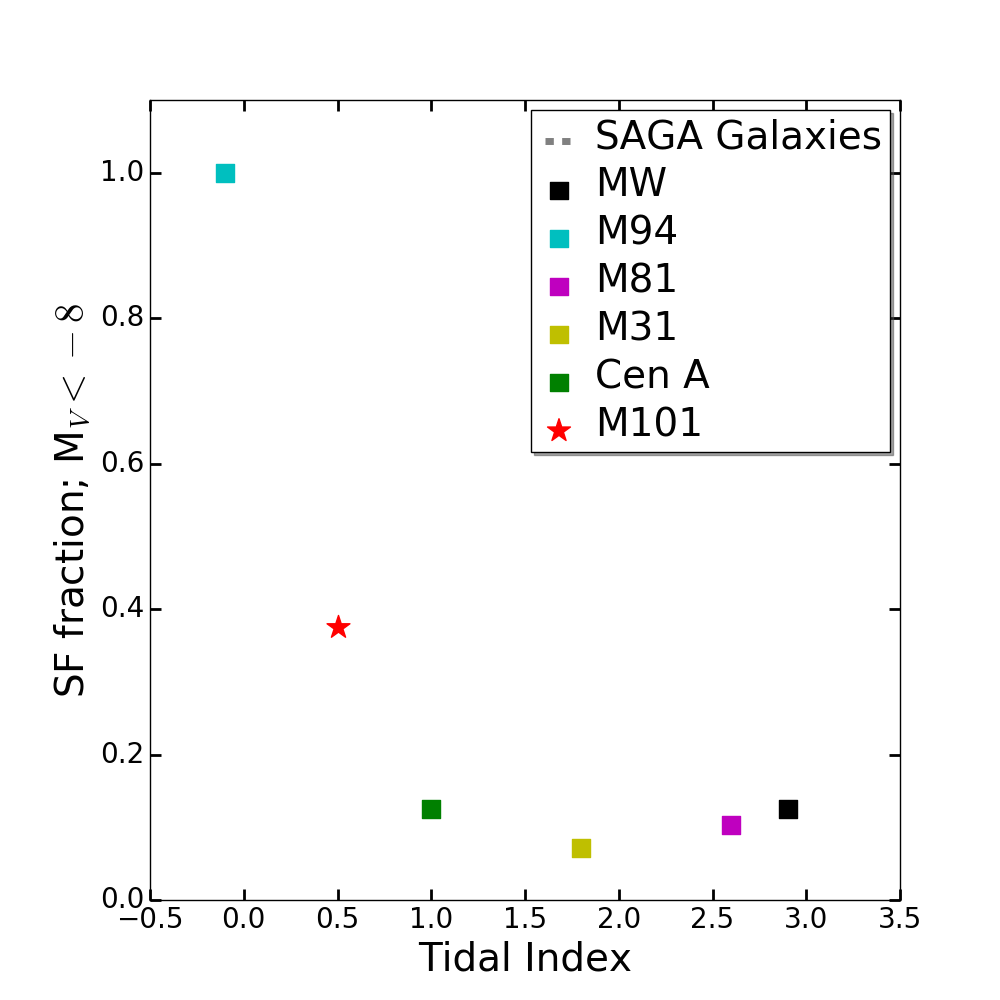}
 \caption{The environmental density of our target galaxies, based on tidal index, against the star forming fraction of satellite galaxies with M$_V$$\leq$$-$12 (left) and M$_V$$\leq$$-$8 (right). M101 is represented by a red star, the other Local Volume galaxies are represented by squares. The gray dashed line represented the average star forming fraction measured by the SAGA survey \citep{Geha17}, whose galaxies do not have reported tidal indices. 
 \label{fig:star_form}}
 \end{center}
\end{figure*}

This relationship between environment and star formation can be examined directly by again using the tidal index parameter \citep{karachentsev13} as described in Section \ref{subsec:density}. In Figure \ref{fig:star_form} we plot the star forming fractions for the satellite galaxies against the tidal index of the host galaxy; we do this for satellites with M$_V$$<$$-$12 and M$_V$$<$$-$8. This figure shows that the objects with low tidal indices have high star forming fractions among their satellites. Again we have also examined a sample only using satellites within 150 kpc projected distance, this yields similar trends where galaxies with larger tidal indices have smaller star forming fractions, however the scatter is larger due to smaller sample sizes.    

The galaxies examined as part of the SAGA survey do not have published tidal indices and these are hard to calculate without full distance information. 
However we can construct strict upper limits, using the projected distance in place of the 3D separation in equation (3). This gives a tidal index for the galaxy if all galaxies were at exactly the same distance. This is a poor assumption and as a result the tidal indecies of these galaxies are likely much lower than this limit. 
These upper limits show that the SAGA galaxies are likely isolated as only 2 of the 8 have upper limits greater than the tidal index of M31 (1.8).  
This provides additional evidence for the conclusion that a lower tidal index in a galaxy leads to a higher star forming fraction in its satellites. 

\subsection{Asymmetry in the Satellite Spatial Distribution}\label{subsec:asymmetry}

An asymmetry among LSB candidates projected around the M101 group was reported in \citet{bennet17}. Our follow-up imaging shows that this asymmetry is caused by the presence of the NGC 5485 group to the northeast of M101, rather than being an innate property of the M101 group. However there is still a curious asymmetry within the M101 group, with the majority of confirmed classical and LSB satellites being found to the southeast. Five of the eight satellites of the M101 group are to the southeast, compared to 2.0$\pm$1.4 that would be expected in a purely random distribution. This asymmetry can be seen in Figure $\ref{fig:pos}$. 
Despite the highly asymmetric HI and optical disk of M101 \citep{mihos12,mihos13} which show extensive features to the northeast and east there are no extensions to the southeast.

Examination of the spatial distribution of the NGC 5485 group, on the other hand, shows no distinct asymmetries aside from a slight overdensity to the south, which is also visible in Figure $\ref{fig:pos}$. This is likely the result of a selection effect, as the area to the south is closer to M101 and has therefore been more widely studied.

\begin{figure}
 \begin{center}
 \includegraphics[width=9cm]{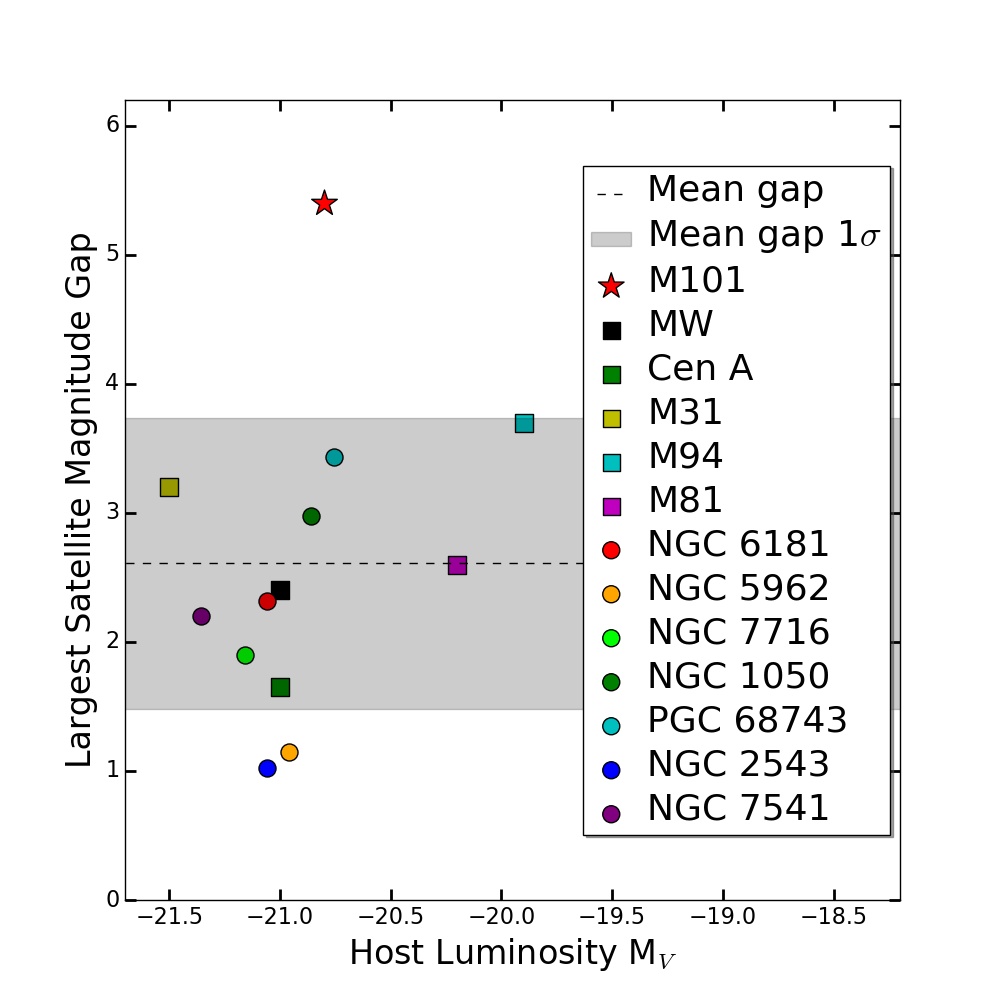}
 \caption{The largest magnitude gap in the satellite luminosity function between two confirmed group members as a function of host magnitude. The largest magnitude gap does not include the gap between the host galaxy and the brightest satellite. The nearby Local Volume galaxy sample from the current work are squares, whereas the SAGA \citep{Geha17} sample is shown as circles, M101 is shown as a red star. The dashed line shows the mean gap size and the shaded area the 1$\sigma$ uncertainty. \label{fig:gap}}
 \end{center}
\end{figure}

\section{Conclusion} \label{sec:conclusion}

In this work we have presented {\it HST} follow-up imaging of two new M101 dwarfs (DwA and Dw9), as well as 17 additional diffuse dwarf candidates which were unresolved and in the background. 
This {\it HST} imaging has allowed us to derive updated values for the distances, luminosities, structural parameters and photometric metallicities for the targeted M101 dwarfs. These values (along with the magnitude and half-light radius of the unresolved candidates) are found to be in broad agreement with previously reported ground based observations from the CFHTLS \citep{bennet17}.  
These new dwarfs have expanded the LF for the M101 group down to M$_{V}$=$-$8.2, on the edge of the ultra-faint dwarf regime. 

Using these new objects we have constructed an updated M101 LF. We have compared this LF to other nearby Local Volume galaxies (MW, M31, M81, M94 and Cen A) and this has shown that M101 has several unusual characteristics. 
The extension of the LF down to M$_V$=$-$8 shows that the M101 group is sparse, with a factor of $\sim$3 fewer satellites than M31, M81 and Cen A. This is highlighted by the fact that M101 has 9 group members brighter than this magnitude compared to the median value of 24.5$\pm$7.7 for other nearby groups (see Figure \ref{fig:LF_NGC}). 
We also find that within the virial radius of M101, there are no confirmed group members in the range $-10>$M$_V>-$15. This means that M101 presents the largest satellite magnitude gap (5.4~mag) that has so far been observed around a MW-mass host: the mean for MW-mass galaxies using both the local sample and the results from the SAGA survey \citep{Geha17} is found to be 2.6$\pm$1.1 mag (see Figure \ref{fig:gap}), with both these values being larger than predictions from simulations. 

Given that M94, another relatively isolated nearby Local Volume galaxy, also hosts significantly fewer satellites than the MW, M31, M81 and Cen A, this may indicate an environmental trend, with the lower density groups showing far fewer satellites than those in denser environments, this relation can be seen in Figure \ref{fig:density}. The observed level of scatter in the LFs between these similarly massive galaxies is larger than can be explained by simulations (see \citealt{Smercina18}). It is clear that further observations of the faint end of satellite LFs for more galaxies are required, in addition to work on simulations to try and reproduce this large scatter and apparent density dependence. 

We have further explored a possible link between a host's galactic environment and star forming fraction within the satellite galaxies. We have shown that groups with tidal index $<$1 seem to have active star formation in all group members with M$_V$$\leq$$-$12, in contrast to denser groups which have star forming fractions of 20-40\% among these group members. This also lines up with results from the SAGA survey \citep{Geha17}, where isolated galaxies were shown to have ongoing star formation in almost all detected satellites. This is shown for satellites with M$_V$$<$$-$12 and M$_V$$<$$-$8 in Figure \ref{fig:star_form}. 

The fact that many of the \cite{bennet17} candidates have been shown to be background objects, despite projection onto the M101 group, should be taken into consideration in future, before drawing conclusions about dwarf populations around nearby galaxies without confirmed distance estimates. The work by \cite{carlsten19} to utilize the SBF distance measurement technique and apply it to LSB galaxies has also proven to be promising. Deeper follow-up observations with {\it HST} are still necessary to constrain the substructure properties of nearby galaxy systems.

\acknowledgments

We are grateful to the referee for a careful reading of the manuscript and for his/her useful suggestions that helped improve this work.

PB and DJS acknowledge useful discussions with B. Weiner during the preparation of this manuscript.

Research by PB is supported by NASA through grant number HST-GO-14796.005-A from the Space Telescope Science Institute which is operated by AURA, Inc., under NASA contract NAS 5-26555. Research by DJS is supported by NSF grants AST-1821967, 1821987, 1813708, 1813466, and 1908972.  Research by DC is supported by NSF grant AST-1814208, and by NASA through grants number HST-GO-15426.007-A and HST-GO-15332.004-A from the Space Telescope Science Institute, which is operated by AURA, Inc., under NASA contract NAS 5-26555.

Based on observations obtained with MegaPrime/MegaCam, a joint project of CFHT and CEA/IRFU, at the Canada-France-Hawaii Telescope (CFHT) which is operated by the National Research Council (NRC) of Canada, the Institut National des Science de l'Univers of the Centre National de la Recherche Scientifique (CNRS) of France, and the University of Hawaii. This work is based in part on data products produced at Terapix available at the Canadian Astronomy Data Centre as part of the Canada-France-Hawaii Telescope Legacy Survey, a collaborative project of NRC and CNRS.
We acknowledge the usage of the HyperLeda database\footnote{http://leda.univ-lyon1.fr}.

\vspace{5mm}
\facilities{Hubble Space Telescope, Canada France Hawaii Telescope (Megacam)}

\software{Astropy \citep{astropy13,astropy18}, SExtractor \citep{bertin96}, GALFIT \citep{peng02}, The IDL Astronomy User's Library \citep{IDLforever}}

\bibliographystyle{aasjournal}
\bibliography{ref_PB}

\end{document}